\documentclass[12pt]{article}
\usepackage{epsfig,amsmath,amsfonts,amssymb,amstext,afterpage,psfrag}
\renewcommand{\arraystretch}{1.2}

\long\def\symbolfootnote[#1]#2{\begingroup%
\def\thefootnote{\fnsymbol{footnote}}\footnote[#1]{#2}\endgroup}

\def\l{\langle}
\def\r{\rangle}

\def\spose#1{\hbox to 0pt{#1\hss}}
\def\lsim{\mathrel{\spose{\lower 3pt\hbox{$\mathchar"218$}}
 \raise 2.0pt\hbox{$\mathchar"13C$}}}
\def\gsim{\mathrel{\spose{\lower 3pt\hbox{$\mathchar"218$}}
 \raise 2.0pt\hbox{$\mathchar"13E$}}}

\catcode`@=11
\def\@citex[#1]#2{%
  \if@filesw\immediate\write\@auxout{\string\citation{#2}}\fi
  \def\@citea{}\@cite{\@for\@citeb:=#2\do
    {\@citea\def\@citea{,\penalty\@m}\@ifundefined
      {b@\@citeb}{{\bf ?}\@warning
{Citation `\@citeb' on page \thepage \space undefined}}%
      \hbox{\csname b@\@citeb\endcsname}}}{#1}}
\def\citer{\@ifnextchar [{\@tempswatrue\@citexr}{\@tempswafalse\@citexr[]}}
  \def\@citexr[#1]#2{%
    \if@filesw\immediate\write\@auxout{\string\citation{#2}}\fi
    \def\@citea{}\@cite{\@for\@citeb:=#2\do
      {\@citea\def\@citea{--\penalty\@m}\@ifundefined
{b@\@citeb}{{\bf ?}\@warning
{Citation `\@citeb' on page \thepage \space undefined}}%
\hbox{\csname b@\@citeb\endcsname}}}{#1}}

\begin{document}

\begin{titlepage}

\begin{flushright}
{\small
LMU-ASC~42/13\\ 
FLAVOUR(267104)-ERC-47\\
July 2013\\
}
\end{flushright}

\vspace{0.5cm}
\begin{center}
{\Large\bf \boldmath                                               
Complete Electroweak Chiral Lagrangian\\      
\vspace*{0.3cm}                                                            
with a Light Higgs at NLO                     
\unboldmath}
\end{center}

\vspace{0.5cm}
\begin{center}
{\sc Gerhard Buchalla, Oscar Cat\`a and Claudius Krause} 
\end{center}

\vspace*{0.4cm}

\begin{center}
Ludwig-Maximilians-Universit\"at M\"unchen, Fakult\"at f\"ur Physik,\\
Arnold Sommerfeld Center for Theoretical Physics, 
D--80333 M\"unchen, Germany
\end{center}

\vspace{1.5cm}
\begin{abstract}
\vspace{0.2cm}\noindent
We consider the Standard Model, including a light scalar boson $h$,
as an effective theory at the weak scale $v=246\,{\rm GeV}$ of
some unknown dynamics of electroweak symmetry breaking. This dynamics
may be strong, with $h$ emerging as a pseudo-Goldstone boson.
The symmetry breaking scale $\Lambda$ is taken to be at $4\pi v$ or above.
We review the leading-order Lagrangian within this framework, which
is nonrenormalizable in general. A chiral Lagrangian can then be constructed
based on a loop expansion. A systematic power counting is derived and
used to identify the classes of counterterms that appear at one loop order.
With this result the complete Lagrangian is constructed at next-to-leading 
order, ${\cal O}(v^2/\Lambda^2)$. This Lagrangian is the most
general effective description of the Standard Model containing
a light scalar boson, in general with strong dynamics of electroweak
symmetry breaking. Scenarios such as the SILH ansatz or
the dimension-6 Lagrangian of a linearly realized Higgs sector
can be recovered as special cases.
\end{abstract}

\vfill

\end{titlepage}

\section{Introduction}
\label{sec:intro}

The recent discovery of a scalar sector in the Standard Model has been one 
of the most important breakthroughs of the last decades in particle physics. 
The additional confirmation, as more and more experimental evidence is piling 
up~\citer{Aad:2012tfa,CMS:yva}, that the scalar particle closely resembles 
the Higgs boson is even more remarkable, meaning that the Standard Model 
provides a rather successful description of electroweak symmetry breaking. 
In particular, recent experimental results strengthen the evidence
for a particle with spin 0 and positive parity \cite{Aad:2013xqa}.

However, the Standard Model solution to electroweak symmetry breaking 
is extremely fine-tuned and should be deemed unsatisfactory. More natural 
solutions typically call for new physics states at the TeV scale, for which 
unfortunately there is no evidence so far. However, their eventual existence 
would typically induce deviations from the Standard Model Higgs parameters, 
which, even if only slight, would be of profound significance for the 
renormalizability and unitarization of the theory and, more generally, for 
our understanding of the dynamics of electroweak symmetry breaking. 

There exists a large number of alternatives to the Higgs model, which 
provide different dynamical explanations of electroweak symmetry breaking. 
From a phenomenological viewpoint it is however more efficient to test 
these potential deviations from the Standard Model with a broader framework 
and then particularize to specific models, the Standard Model being one of 
them. Given the large energy gap between the electroweak scale 
$v=246$~GeV and the expected new physics scale $\Lambda\sim$~few TeV, this 
broader framework can be most easily cast in an effective field theory 
(EFT) language. This EFT should provide, by construction, the most general 
description of the electroweak interactions in the presence of a light 
scalar $h$, and therefore provide the right framework to test its dynamical 
nature. As a result, the EFT we are after is actually the most general EFT 
description of the electroweak interactions with the presently known 
particle content. 

The starting point for such an EFT requires a parameterization of the minimal 
coset $SU(2)_L\times U(1)_Y/U(1)_{em}$, which can be done using a nonlinear 
realization~\cite{Coleman:1969sm}. The resulting Goldstone bosons provide the 
longitudinal components of the gauge bosons. The new scalar $h$ is then 
introduced in full generality as a singlet under $SU(2)_L\times U(1)_Y$. This 
path has been pursued before, and partial sets of the resulting 
effective-theory operators have been listed and their 
phenomenological consequences explored~\citer{Feruglio:1992wf,Azatov:2012bz}. 

However, the previous papers 
lacked a careful discussion of the foundations of the EFT, including 
essential aspects in the construction of the operator basis such as 
power-counting arguments. In this paper we want to fill this gap and put the 
EFT on a more systematic basis. A large part of this effort was already done 
in \cite{Buchalla:2012qq}, where the systematics of the nonlinear EFT of 
electroweak interactions was spelled out. In this paper we show how to 
extend those results when a scalar singlet $h$ is included. 

This paper is organized as follows: in Section~\ref{sec:lchilo} we review 
the Standard Model chiral Lagrangian at leading order as the most general 
description of electroweak symmetry breaking. In Section~\ref{sec:powc} 
we discuss how to organize the EFT expansion in powers of $v^2/\Lambda^2$ with 
a consistent power-counting. Section~\ref{sec:lhnlo} is devoted to working 
out the most general basis of operators at next-to-leading order (NLO). 
In Section~\ref{sec:xicount} we extend our discussion to include generic 
scenarios of partial compositeness as interpolations between the 
purely strongly-coupled and weakly-coupled limits. A comparison with the 
previous literature is provided in Section~\ref{sec:complit}. For illustration, 
in Section~\ref{sec:uvmod} we include two particular model realizations, 
namely the $SO(5)/SO(4)$ composite Higgs model and a Higgs-portal model, 
showing how they reduce to particular parameter choices of the 
general EFT. Conclusions are given in Section~\ref{sec:concl}, while 
technical details are collected in an Appendix.


\section{SM chiral Lagrangian at leading order}
\label{sec:lchilo}

In this section we summarize the leading-order (LO) electroweak chiral 
Lagrangian of the Standard Model including a light Higgs field $h$.
Further comments on the systematics behind its construction can be found
in Appendix~\ref{sec:appllo}. 

The leading-order Lagrangian can be written as
\begin{equation}\label{l4uh}
{\cal L}_{LO} = {\cal L}_4 + {\cal L}_{Uh}
\end{equation} 
The first term, ${\cal L}_4$, represents the unbroken, renormalizable
part, built from the left-handed doublets $q$, $l$ and right-handed 
singlets $u$, $d$, $e$ of quarks and leptons, together with the gauge fields 
$G$, $W$, $B$ of $SU(3)_C$, $SU(2)_L$, $U(1)_Y$: 
\begin{eqnarray}\label{lsm4}
{\cal L}_4 &=& -\frac{1}{2} \langle G_{\mu\nu}G^{\mu\nu}\rangle
-\frac{1}{2}\langle W_{\mu\nu}W^{\mu\nu}\rangle 
-\frac{1}{4} B_{\mu\nu}B^{\mu\nu}
\nonumber\\
&& +\bar q i\!\not\!\! Dq +\bar l i\!\not\!\! Dl
 +\bar u i\!\not\!\! Du +\bar d i\!\not\!\! Dd +\bar e i\!\not\!\! De 
\end{eqnarray}
Generation indices have been omitted.
Here and in the following the trace of a matrix $M$ is denoted by
$\langle M\rangle$.
The covariant derivative of a fermion field $\psi_{L,R}$ is defined as
\begin{equation}\label{dcovf}
D_\mu\psi_L =
\partial_\mu \psi_L +i g W_\mu \psi_L +i g' Y_{\psi_L} B_\mu \psi_L , 
\qquad
D_\mu\psi_R =\partial_\mu \psi_R  + i g' Y_{\psi_R} B_\mu \psi_R
\end{equation}
dropping the QCD part for simplicity. The Higgs-sector Lagrangian reads
\begin{eqnarray}\label{luh}
{\cal L}_{Uh} &=& 
\frac{v^2}{4}\ \l D_\mu U^\dagger D^\mu U\r\, \left( 1+F_U(h)\right)
+\frac{1}{2} \partial_\mu h \partial^\mu h - V(h) \nonumber\\
&& - v \left[ \bar q \left(\hat Y_u +
         \sum^\infty_{n=1}\hat Y^{(n)}_u \left(\frac{h}{v}\right)^n \right) U P_+r 
+ \bar q \left(\hat Y_d + 
         \sum^\infty_{n=1}\hat Y^{(n)}_d \left(\frac{h}{v}\right)^n \right) U P_-r
  \right. \nonumber\\ 
&& \quad\quad\left. + \bar l \left(\hat Y_e +
   \sum^\infty_{n=1}\hat Y^{(n)}_e \left(\frac{h}{v}\right)^n \right) U P_-\eta 
+ {\rm h.c.}\right]
\end{eqnarray}
where
\begin{equation}\label{fuvsum}
F_U(h)=\sum^\infty_{n=1}f_{U,n}\left(\frac{h}{v}\right)^n\, ,\qquad
V(h)=v^4 \sum^\infty_{n=2}f_{V,n}\left(\frac{h}{v}\right)^n
\end{equation}
Here the right-handed quark and lepton fields are written as
$r=(u,d)^T$ and $\eta=(\nu,e)^T$, respectively. In general,
different flavour couplings $\hat Y^{(n)}_{u,d,e}$ can arise at every order in 
the Higgs field $h^n$, in addition to the usual Yukawa matrices $\hat Y_{u,d,e}$.
We define 
\begin{equation}\label{pm12def}
P_\pm\equiv \frac{1}{2}\pm T_3\, ,\qquad 
P_{12}\equiv T_1+i T_2\, ,\qquad P_{21}\equiv T_1-i T_2
\end{equation}
where $P_{12}$ and $P_{21}$ will be needed later on. 

Under $SU(2)_L\times SU(2)_R$ the Goldstone boson
matrix $U$ and the Higgs-singlet field $h$ transform as
\begin{equation}\label{uhglgr}
U\rightarrow g_L U g^\dagger_R,\qquad h\rightarrow h,
\qquad g_{L,R}\in SU(2)_{L,R}
\end{equation}
The transformations $g_L$ and the $U(1)_Y$ subgroup of $g_R$
are gauged, so that the covariant derivatives are given by
\begin{equation}\label{dcovu}
D_\mu U=\partial_\mu U+i g W_\mu U -i g' B_\mu U T_3,
\qquad D_\mu h = \partial_\mu h
\end{equation}
The explicit relation between the matrix $U$ and the Goldstone fields
$\varphi^a$ is
\begin{equation}\label{uudef}
U=\exp(2i\Phi/v),\qquad
\Phi=\varphi^a T^a=\frac{1}{\sqrt{2}}\left(
\begin{array}{cc}
\frac{\varphi^0}{\sqrt{2}} & \varphi^+\\
\varphi^- & -\frac{\varphi^0}{\sqrt{2}} 
\end{array}\right)
\end{equation}
where $T^a=T_a$ are the generators of $SU(2)$.

\section{Power counting}
\label{sec:powc}

The leading-order Lagrangian (\ref{l4uh}) is nonrenormalizable
in general. A consistent effective field theory can be constructed
order by order in the loop expansion. The next-to-leading order 
terms can be classified according to the counterterms that appear
at one loop. The corresponding classes of operators are determined
by standard methods of power counting. For the case of the chiral
Lagrangian in (\ref{l4uh}) without the Higgs scalar $h$ this
procedure has been discussed in \cite{Buchalla:2012qq}, where further
details can be found.
The generalization to include the light Higgs scalar is straightforward
and will be summarized in the following. We will omit ghost fields, which 
insure manifest gauge independence, but do not affect the power counting.  

Without $h$, a generic $L$-loop diagram ${\cal D}$, built from
(\ref{l4uh}), contains  
$n_i$ $\varphi^{2i}$-vertices and $\nu_k$ Yukawa interactions 
$\bar\psi_{L(R)}\psi_{R(L)}\varphi^k$, a number $m_l$ of gauge-boson-Goldstone  
vertices $X_\mu \varphi^l$, $r_s$ such vertices of the type
$X^2_\mu  \varphi^s$, $x$ quartic gauge-boson vertices $X^4_\mu$, 
$u$ triple-gauge-boson vertices $X^3_\mu$, and $z_L$ ($z_R$)
fermion-gauge-boson interactions $\bar\psi_{L(R)}\psi_{L(R)}X_\mu$. 
Here $\psi_L$ ($\psi_R$), $\varphi$ and $X_\mu$ denote
left-handed (right-handed) fermions, Goldstone bosons and
gauge fields, respectively.

The presence of $h$ introduces into ${\cal D}$ a number $\sigma_{ja}$ of 
Goldstone-Higgs vertices $\varphi^{2j} h^a$, $\tau_{tb}$ Yukawa
vertices with $t$ Goldstone and $b$ Higgs lines, as well as
$\omega_q$ $h^q$-interactions.

Following the steps discussed in \cite{Buchalla:2012qq}, the power-counting 
for the diagram ${\cal D}$ can be summarized by the formula 
\begin{equation}\label{pcupsix}
{\cal D}\sim 
\frac{(yv)^\nu (gv)^{m+2r+2x+u+z}}{v^{F_L+F_R-2-2\omega}} \frac{p^d}{\Lambda^{2L}}\
\bar\psi_L^{F^1_L} \psi_L^{F^2_L} \bar\psi_R^{F^1_R} \psi_R^{F^2_R}\
\left(\frac{X_{\mu\nu}}{v}\right)^V\  \left(\frac{\varphi}{v}\right)^B\
\left(\frac{h}{v}\right)^H
\end{equation}
where the power of external momenta $p$ is
\begin{equation}\label{powerd}
d\equiv 2L+2-\frac{F_L+F_R}{2}-V-\nu-m-2r-2x-u-z-2\omega
\end{equation}
Here $F_L=F^1_L+F^2_L$, $F_R=F^1_R+F^2_R$ and $V$  
is the number of external left-handed fermion, right-handed fermion 
and gauge-boson lines, respectively. $g$ is a generic 
gauge coupling, and we have used $\nu\equiv\sum_k\nu_k+\sum_{t,b}\tau_{tb}$, 
$m\equiv\sum_l m_l$, $r\equiv\sum_s r_s$, $z\equiv z_L+z_R$,
$\omega\equiv\sum_q \omega_q$.
An exponent $d\geq 0$ in (\ref{pcupsix}) indicates a divergence by
power counting, as well as the number of derivatives in the corresponding
counterterm.
The expression (\ref{powerd}) for $d$ is useful, because
$F_L$, $F_R$ and $V$, as well as the numbers of vertices, all enter 
with a negative sign. This implies that the number of
divergent diagrams at a given order in $L$ is finite.
We also note that the numbers of both external Goldstone and Higgs boson lines,
$B$ and $H$, enter the power counting formula (\ref{pcupsix}) only
through the factors $(\varphi/v)^B$ and $(h/v)^H$. 
They are irrelevant in particular for the 
exponent $d$, which counts the powers of momentum. This indicates explicitly
that, at any given order in the effective theory, the counterterms
contain an arbitrary number of Goldstone fields $U=U(\varphi/v)$, as well as
Higgs fields $h/v$. Both $\varphi$ and $h$ are therefore on the same footing.
This result of power counting is in agreement with the discussion in
Appendix \ref{sec:appllo}.

Since (\ref{pcupsix}) and (\ref{powerd}) are very similar to the
case without $h$ discussed in \cite{Buchalla:2012qq}, the
generalization to the scenario that includes $h$ follows immediately.
The NLO counterterms are found by enumerating the classes
of diagrams that give rise to a degree of divergence $d\geq 0$
with $L=1$ in (\ref{powerd}). 
Denoting by $Uh$ the presence of any number of Goldstone fields
$U$ (or $U^\dagger$) and Higgs singlets $h$, 
and by $D^n$, $\psi^F$, $X^k$ the numbers
$n$, $F$, $k$, respectively, of derivatives, fermion fields and gauge-boson
field-strength tensors, these classes are schematically given by
\begin{equation}\label{nloclass}                                         
UhD^4,\qquad X^2 Uh,\qquad XUhD^2,\qquad \psi^2 UhD,\qquad \psi^2 UhD^2,\qquad   
\psi^4 Uh                                                              
\end{equation}
The next section will be devoted to constructing the full set of basis
operators in each class.

\section{Effective Lagrangian at next-to-leading order}
\label{sec:lhnlo}

The NLO operators are conveniently expressed using the definitions 
\begin{equation}\label{ltauldef}
L_\mu\equiv i U D_\mu U^\dagger\, , \qquad \tau_L\equiv U T_3 U^\dagger
\end{equation}
Both $L_\mu$ and $\tau_L$ are hermitean and traceless.
They obey the identities
\begin{equation}\label{dmulnu}
D_\mu L_\nu - D_\nu L_\mu = g W_{\mu\nu}- g' B_{\mu\nu}\tau_L + i[L_\mu, L_\nu]
\end{equation}
\begin{equation}\label{dmutaul}
D_\mu \tau_L = - i[\tau_L, L_\mu]
\end{equation}
 
The NLO operators can be constructed using elementary building blocks,
as reviewed in \cite{Buchalla:2012qq} for the case without $h$ field.
In the Goldstone-Higgs sector the required building blocks are
\begin{equation}\label{bbuh}
\langle L_\mu L_\nu\rangle ,\quad \langle \tau_L L_\mu\rangle ,\quad
\langle L_\mu L_\nu L_\lambda\rangle ,\quad \langle\tau_L L_\mu L_\nu\rangle ,
\quad \partial_\mu h ,\quad F(h)
\end{equation}
where $F(h)$ denotes a generic function of $h/v$.
Five additional building blocks arise when the electroweak field strengths
are included
\begin{equation}\label{bbxu}
\langle W_{\mu\nu} L_\lambda\rangle ,\quad\langle\tau_L W_{\mu\nu}\rangle ,\quad
\langle W_{\mu\nu} L_\lambda L_\rho\rangle ,\quad
\langle \tau_L W_{\mu\nu} L_\lambda\rangle ,\quad B_{\mu\nu}
\end{equation}
Together with the terms in the LO Lagrangian, these elements are sufficient 
to construct the NLO operators in the purely bosonic sector. Operators with 
fermions can be obtained along similar lines \cite{Buchalla:2012qq}. 
Note that apart from the functions $F(h)$, which enter each operator 
as an overall factor, the only new building block in comparison 
to \cite{Buchalla:2012qq} is $\partial_\mu h$.

Using integration by parts, the identities (\ref{dmulnu}) and (\ref{dmutaul}), 
and the leading-order equations of motion, certain operators can be
shown to be redundant. To proceed in a systematic way, we eliminate a given
operator, if possible, in favour of operators with fewer derivatives.

The next-to-leading-order effective Lagrangian of the Standard Model 
with dynamically broken electroweak symmetry, including a light Higgs scalar, 
can then be written as
\begin{equation}\label{leffnlo}
{\cal L}={\cal L}_{LO} + {\cal L}_{\beta_1} +
\sum_i c_i\frac{v^{6-d_i}}{\Lambda^2}\, {\cal O}_i
\end{equation}
Here ${\cal L}_{LO}$ is the leading order Lagrangian (\ref{l4uh})
and ${\cal L}_{\beta_1}$ the custodial-symmetry breaking, 
dimension-2 operator 
\begin{equation}\label{lbeta1}
{\cal L}_{\beta_1} = 
-\beta_1 v^2 \langle\tau_L L_\mu\rangle\, \langle\tau_L L^\mu\rangle\, 
F_{\beta_1}(h),\qquad  
F_{\beta_1}(h)=1+\sum_{n=1}^\infty f_{\beta_1,n}\left(\frac{h}{v}\right)^n
\end{equation}
As discussed in Appendix~\ref{sec:appllo} this operator can be treated 
as a next-to-leading order correction. Apart from this term, the 
NLO operators are denoted by ${\cal O}_i$ in (\ref{leffnlo}). 
They come with a suppression by two powers of the symmetry-breaking scale 
$\Lambda\approx 4\pi v$ and have dimensionless coefficients $c_i$, 
which are naturally of order unity. 
$d_i$ is the canonical dimension of the operator ${\cal O}_i$.
Conservation of baryon and lepton number will be assumed in the 
present context, since their violation is expected to arise only
at scales much above the few TeV range. 
Further remarks can be found in \cite{Buchalla:2012qq}.

In the following we list the NLO operators ${\cal O}_i$
according to the classification introduced at the end of 
Section \ref{sec:powc}.

\subsection{\boldmath $UhD^4$ terms}
\label{subsec:uhd4}

The operators of this class generalize the ${\cal O}(p^4)$ 
chiral-Lagrangian terms $UD^4$ already given in \cite{Longhitano:1980tm},
now including arbitrary powers of $h/v$.
There is a total of 15 independent operators, of which 11 are CP even
and 4 are CP odd. 

The CP even operators can be written as
\begin{eqnarray}\label{u4h0}  
{\cal O}_{D1} &=& \l L_\mu L^\mu \r^2 \ F_{D1}(h)\nonumber\\
{\cal O}_{D2} &=& \l L_\mu L_\nu\r \ \l L^\mu L^\nu\r  \ F_{D2}(h)\nonumber\\
{\cal O}_{D3} &=& \left(\l\tau_L L_\mu\r\ 
       \l\tau_L L^\mu\r\right)^2 \ F_{D3}(h)\nonumber\\ 
{\cal O}_{D4} &=& \l \tau_L L_\mu \r\ \l\tau_L L^\mu\r 
  \ \l L_\nu L^\nu \r \ F_{D4}(h)\nonumber\\
{\cal O}_{D5} &=& \l\tau_L L_\mu\r\  \l\tau_L L_\nu\r \ \l L^\mu L^\nu\r\ F_{D5}(h)
\end{eqnarray}

\begin{eqnarray}\label{u3h1}
{\cal O}_{D6} &=& i \l\tau_L L_\mu L_\nu \r \ \l\tau_L L^\mu\r\ 
\frac{\partial^\nu h}{v} F_{D6}(h)
\end{eqnarray}

\begin{eqnarray}\label{u2h2}
{\cal O}_{D7} &=& \l L_\mu L^\mu \r\ 
  \ \frac{\partial_\nu h\, \partial^\nu h}{v^2} F_{D7}(h)\nonumber\\
{\cal O}_{D8} &=& \l L_\mu L_\nu \r\ 
  \ \frac{\partial^\mu h\, \partial^\nu h}{v^2} F_{D8}(h)\nonumber\\
{\cal O}_{D9} &=& \l\tau_L L_\mu \r\  \l\tau_L L^\mu \r 
  \ \frac{\partial_\nu h\, \partial^\nu h}{v^2} F_{D9}(h)\nonumber\\
{\cal O}_{D10} &=& \l\tau_L L_\mu \r\  \l\tau_L L_\nu \r 
  \ \frac{\partial^\mu h\, \partial^\nu h}{v^2} F_{D10}(h)
\end{eqnarray}

\begin{eqnarray}\label{u0h4}
{\cal O}_{D11} &=& \frac{(\partial_\mu h\, \partial^\mu h)^2}{v^4} F_{D11}(h)
\end{eqnarray}
The CP odd operators are
\begin{eqnarray}\label{uhodd}
{\cal O}_{D12} &=& \l L_\mu L^\mu \r\ 
  \l\tau_L L_\nu \r \ \frac{\partial^\nu h}{v} F_{D12}(h) \nonumber\\
{\cal O}_{D13} &=& \l L_\mu L_\nu \r \ \l\tau_L L^\mu\r\ 
\frac{\partial^\nu h}{v} F_{D13}(h) \nonumber\\
{\cal O}_{D14} &=& \l\tau_L L_\mu \r\  \l\tau_L L^\mu \r\ 
  \l\tau_L L_\nu \r \ \frac{\partial^\nu h}{v} F_{D14}(h) \nonumber\\
{\cal O}_{D15} &=& \l\tau_L L_\mu \r
  \ \frac{\partial^\mu h\, \partial_\nu h\, \partial^\nu h}{v^3} F_{D15}(h)
\end{eqnarray}
We have defined
\begin{equation}
F_{Di}(h)\equiv 1+ \sum_{n=1}^\infty f_{Di,n}\left(\frac{h}{v}\right)^n
\end{equation}

The four subclasses in (\ref{u4h0}), (\ref{u3h1}), (\ref{u2h2})  
and (\ref{u0h4}) correspond, respectively, to terms
with zero, one, two and four derivatives acting on $h$.  
The subclass of CP odd operators has terms with one derivative
acting on $h$ and contains the only operator with three derivatives
on $h$. Note that all operators are written with only single derivatives
on either $U$ or $h$ fields. 
In the absence of the field $h$ the basis
reduces to the five operators in (\ref{u4h0}) with $F_{Di}=1$,
known from \cite{Longhitano:1980tm}.

If custodial symmetry is respected by the $UhD^4$ terms, the basis
reduces to the five operators ${\cal O}_{Di}$ with 
$i=1$,~$2$,~$7$,~$8$ and $11$, all of which are CP even.  
The custodial-symmetry violating $UhD^4$ operators are not generated as
one-loop counterterms if the leading-order Goldstone-Higgs sector is
custodial symmetric. They might still appear as finite contributions at NLO.

\subsection{\boldmath $X^2Uh$ and $XUhD^2$ terms}
\label{subsec:xuhd}

The CP-even operators are 
\begin{align}\label{xhev}
{\cal{O}}_{Xh1}&=g^{\prime 2} B_{\mu\nu} B^{\mu\nu} \, F_{Xh1}(h)\nonumber\\
{\cal{O}}_{Xh2}&=g^2 \langle W_{\mu\nu} W^{\mu\nu}\rangle \, F_{Xh2}(h)\nonumber\\
{\cal{O}}_{Xh3}&=g^2_s \langle G_{\mu\nu} G^{\mu\nu}\rangle \, F_{Xh3}(h)
\end{align}
\begin{align}\label{xuev}
{\cal{O}}_{XU1}&=g^{\prime}gB_{\mu\nu}\langle W^{\mu\nu}\tau_L\rangle
\, (1+F_{XU1}(h))\nonumber\\
{\cal{O}}_{XU2}&=g^2 \langle W_{\mu\nu}\tau_L\rangle^2 \, (1+F_{XU2}(h))\nonumber\\
{\cal{O}}_{XU3}&=g\varepsilon_{\mu\nu\lambda\rho}
\langle W^{\mu\nu}L^{\lambda}\rangle\langle\tau_L L^{\rho}\rangle
\, (1+F_{XU3}(h))\nonumber\\
{\cal{O}}_{XU7}&=ig^{\prime}B_{\mu\nu}\langle\tau_L[L^{\mu},L^{\nu}]
\rangle \, F_{XU7}(h)\nonumber\\
{\cal{O}}_{XU8}&=ig\langle W_{\mu\nu}[L^{\mu},L^{\nu}]
\rangle \, F_{XU8}(h)\nonumber\\
{\cal{O}}_{XU9}&=ig\langle W_{\mu\nu}\tau_L\rangle 
\langle \tau_L[L^{\mu},L^{\nu}]\rangle \, F_{XU9}(h)
\end{align}
In correspondence to (\ref{xhev}) and (\ref{xuev}) 
there are also nine CP-odd operators:
\begin{align}\label{xhod}
{\cal{O}}_{Xh4}&=g^{\prime 2} \varepsilon_{\mu\nu\lambda\rho}
B^{\mu\nu} B^{\lambda\rho} \, F_{Xh4}(h)\nonumber\\
{\cal{O}}_{Xh5}&=g^2 \varepsilon_{\mu\nu\lambda\rho}
\langle W^{\mu\nu} W^{\lambda\rho}\rangle \, F_{Xh5}(h)\nonumber\\
{\cal{O}}_{Xh6}&=g^2_s \varepsilon_{\mu\nu\lambda\rho}
\langle G^{\mu\nu} G^{\lambda\rho}\rangle \, F_{Xh6}(h)
\end{align}
\begin{align}\label{xuod}
{\cal{O}}_{XU4}
&=g^{\prime}g\varepsilon_{\mu\nu\lambda\rho}\langle \tau_L W^{\mu\nu}\rangle
B^{\lambda\rho}\, (1+F_{XU4}(h))\nonumber\\
{\cal{O}}_{XU5}&=g^2\varepsilon_{\mu\nu\lambda\rho}\langle\tau_LW^{\mu\nu}\rangle
\langle\tau_LW^{\lambda\rho}\rangle \, (1+F_{XU5}(h)) \nonumber\\
{\cal{O}}_{XU6}&=g\langle W_{\mu\nu}L^{\mu}\rangle\langle \tau_LL^{\nu}\rangle \,
(1+F_{XU6}(h))\nonumber\\
{\cal{O}}_{XU10}
&=ig^{\prime}\varepsilon_{\mu\nu\lambda\rho}B^{\mu\nu}\langle\tau_L[L^{\lambda},L^{\rho}]
\rangle \, F_{XU10}(h)\nonumber\\
{\cal{O}}_{XU11}
&=ig\varepsilon_{\mu\nu\lambda\rho}\langle W^{\mu\nu}[L^{\lambda},L^{\rho}]
\rangle \, F_{XU11}(h)\nonumber\\
{\cal{O}}_{XU12}&=ig\varepsilon_{\mu\nu\lambda\rho}\langle W^{\mu\nu}\tau_L\rangle 
\langle \tau_L[L^{\lambda},L^{\rho}]
\rangle \, F_{XU12}(h)
\end{align}
Here
\begin{equation}\label{fxuh}
F_{Xi}(h)=\sum^\infty_{n=1}f_{Xi,n}\left(\frac{h}{v}\right)^n
\end{equation}
The terms ${\cal{O}}_{XUi}$, $i=1,\ldots , 6$, remain
independent operators in the limit $h\to 0$, while all other
operators become redundant. For this reason the former operators
are multiplied by $(1+F_{Xi}(h))$.
Omitting the functions $F_{Xi}$, the operators ${\cal O}_{XUi}$ reduce to
those listed already in \cite{Longhitano:1980tm,Appelquist:1993ka}.

\subsection{\boldmath $\psi^2UhD$ terms}

The operators in this class are given by
\begin{equation}\label{psi2uhd}
\begin{array}{cc}
{\cal O}_{\psi V1}=-\bar q\gamma^\mu q\ \l \tau_L L_\mu \r\, F_{\psi V1}(h)
\quad &
{\cal O}_{\psi V4}=-\bar u\gamma^\mu u\ \l \tau_L L_\mu \r\, F_{\psi V4}(h) 
\vspace*{0.2cm}\\
{\cal O}_{\psi V2}=-\bar q\gamma^\mu \tau_L q\ \l\tau_L L_\mu \r\, F_{\psi V2}(h)
\quad &
{\cal O}_{\psi V5}=-\bar d\gamma^\mu d\ \l\tau_L L_\mu \r\, F_{\psi V5}(h) 
\vspace*{0.2cm}\\
{\cal O}_{\psi V3}=-\bar q\gamma^\mu U P_{12} U^\dagger q\ 
\l L_\mu U P_{21}U^\dagger\r\, F_{\psi V3}(h) \quad &
{\cal O}_{\psi V6}=-\bar u\gamma^\mu d\ 
\l L_\mu UP_{21} U^\dagger\r\,  F_{\psi V6}(h) 
\vspace*{0.2cm}\\
{\cal O}^\dagger_{\psi V3} \quad & {\cal O}^\dagger_{\psi V6} \\
& \\
{\cal O}_{\psi V7}=-\bar l\gamma^\mu l\ \l\tau_L L_\mu \r\, F_{\psi V7}(h)
\quad &
{\cal O}_{\psi V10}=-\bar e\gamma^\mu e\ \l\tau_L L_\mu \r\, F_{\psi V10}(h)  
\vspace*{0.2cm}\\
{\cal O}_{\psi V8}=-\bar l\gamma^\mu \tau_L  l\  \l\tau_L L_\mu \r\, F_{\psi V8}(h)  
\quad & 
\vspace*{0.2cm}\\
{\cal O}_{\psi V9}=-\bar l\gamma^\mu U P_{12} U^\dagger l\ 
\l L_\mu U P_{21} U^\dagger\r\, F_{\psi V9}(h) \quad &
{\cal O}^\dagger_{\psi V9} 
\end{array}
\vspace*{0.2cm}
\end{equation}
where
\begin{equation}\label{fvuh}
F_{\psi Vi}(h)=1+\sum^\infty_{n=1}f_{\psi Vi,n}\left(\frac{h}{v}\right)^n
\end{equation}
They generalize the terms first listed for the case without $h$ 
in \cite{Appelquist:1984rr}.
The minus signs on the r.h.s. of (\ref{psi2uhd}) have
been introduced to be consistent with the notation of \cite{Buchalla:2012qq}
in the limit $F_{\psi Vi}(h)\to 1$.
In the sector with left-handed quarks $q$, the four operators
${\cal O}_{\psi V1}$, ${\cal O}_{\psi V2}$, ${\cal O}_{\psi V3}$ and 
${\cal O}^\dagger_{\psi V3}$ are equivalent to the four terms
$\bar q\gamma^\mu q \langle\tau_L L_\mu\rangle F$,
$\bar q\gamma^\mu \tau_L q \langle\tau_L L_\mu\rangle F$, 
$\bar q\gamma^\mu L_\mu q F$ and $\bar q\gamma^\mu i[\tau_L,L_\mu]q F$,
obtained as the independent structures formed directly with 
the building blocks $\tau_L$, $L_\mu$ and a (generic) $F(h)$.
We prefer to work with ${\cal O}_{\psi V3}$ and ${\cal O}^\dagger_{\psi V3}$
in (\ref{psi2uhd}) since in unitary gauge these operators simply 
correspond to charged-current interactions with $W^\pm$.
Taking into account the remaining building block $\partial_\mu h$,
two further operators may be written down,
$\bar q\gamma^\mu q \partial_\mu h F$ and
$\bar q\gamma^\mu \tau_L q \partial_\mu h F$. These are seen to be redundant 
upon integrating by parts, and using the fermion equations of motion and
the identity in (\ref{dmutaul}). Similar comments apply to the operators with 
right-handed quarks and with leptons. The operators in class $\psi^2UhD$ are 
therefore identical to those in class $\psi^2UD$ of \cite{Buchalla:2012qq},
up to overall factors of $F(h)$.

\subsection{\boldmath $\psi^2UhD^2$ and $\psi^2UhX$ terms}

The class $\psi^{2}UhD^{2}$ contains fermion bilinears with Lorentz-scalar
or tensor structure.
The scalar operators are (hermitean conjugate versions will not be
listed separately in this section) 
\begin{equation}
\begin{array}{ll}
\label{psi2uhd2s}
\mathcal{O}_{\psi S1} = \bar{q} U P_{+}r \langle L_{\mu}L^{\mu}\rangle 
F_{\psi S1} 
&\mathcal{O}_{\psi S10} = \bar{q} U P_{+}r \langle\tau_{L}L_{\mu}\rangle 
\left(\partial^{\mu}\tfrac{h}{v}\right) F_{\psi S10}\\ 
\mathcal{O}_{\psi S2} = \bar{q} U P_{-}r \langle L_{\mu}L^{\mu}\rangle F_{\psi S2} 
&\mathcal{O}_{\psi S11} = \bar{q} U P_{-}r \langle\tau_{L}L_{\mu}\rangle 
\left(\partial^\mu \tfrac{h}{v}\right) F_{\psi S11} \\
\rule{0cm}{0.8cm}
\mathcal{O}_{\psi S3} = 
\bar{q} U P_{+}r \langle\tau_{L}L_{\mu}\rangle \langle\tau_{L}L^{\mu}\rangle 
F_{\psi S3}
&\mathcal{O}_{\psi S12} = \bar{q} U P_{12}r 
\langle U P_{21}U^{\dagger}L_{\mu}\rangle\left(\partial^{\mu}\tfrac{h}{v}\right)
F_{\psi S12}\\
\mathcal{O}_{\psi S4} = 
\bar{q} U P_{-}r \langle\tau_{L}L_{\mu}\rangle \langle\tau_{L}L^{\mu}\rangle 
F_{\psi S4}
&\mathcal{O}_{\psi S13} = \bar{q} U P_{21}r 
\langle U P_{12}U^{\dagger}L_{\mu}\rangle\left(\partial^{\mu}\tfrac{h}{v}\right)
F_{\psi S13}\\ 
\rule{0cm}{0.8cm}
\mathcal{O}_{\psi S5} = \bar{q} U P_{12}r\langle\tau_{L}L_{\mu}\rangle 
\langle U P_{21}U^{\dagger}L^{\mu}\rangle F_{\psi S5} 
&\mathcal{O}_{\psi S14} = \bar{q} U P_{+}r 
\left(\partial_{\mu}\tfrac{h}{v}\right)\left(\partial^{\mu}\tfrac{h}{v}\right)
F_{\psi S14} \\
\mathcal{O}_{\psi S6} = \bar{q} U P_{21}r\langle\tau_{L}L_{\mu}\rangle 
\langle U P_{12}U^{\dagger}L^{\mu}\rangle F_{\psi S6} 
&\mathcal{O}_{\psi S15} = \bar{q} U P_{-}r 
\left(\partial_{\mu}\tfrac{h}{v}\right)\left(\partial^{\mu}\tfrac{h}{v}\right)
F_{\psi S15}\\ 
\rule{0cm}{0.8cm}
\mathcal{O}_{\psi S7} =\bar{l} U P_{-}\eta \langle L_{\mu}L^{\mu}\rangle 
F_{\psi S7}
&\mathcal{O}_{\psi S16} = \bar{l} U P_{-}\eta \langle\tau_{L}L_{\mu}\rangle 
\left(\partial^{\mu}\tfrac{h}{v}\right) F_{\psi S16}\\
\mathcal{O}_{\psi S8} = \bar{l} U P_{-}\eta \langle\tau_{L}L_{\mu}\rangle 
\langle\tau_{L}L^{\mu}\rangle F_{\psi S8}
&\mathcal{O}_{\psi S17} = \bar{l} U P_{12}\eta 
\langle U P_{21}U^{\dagger}L_{\mu}\rangle\left(\partial^{\mu}\tfrac{h}{v}\right)
F_{\psi S17}\\
\rule{0cm}{0.8cm}
\mathcal{O}_{\psi S9} = \bar{l} U P_{12}\eta \langle\tau_{L}L_{\mu}\rangle 
\langle U P_{21}U^{\dagger}L^{\mu}\rangle F_{\psi S9} 
&\mathcal{O}_{\psi S18} = \bar{l} U P_{-}\eta 
\left(\partial_{\mu}\tfrac{h}{v}\right)\left(\partial^{\mu}\tfrac{h}{v}\right)
F_{\psi S18}
\end{array}
\end{equation}
The list of operators with a tensor current is
\begin{equation}
\begin{array}{ll}
\label{psi2uhd2t}
\mathcal{O}_{\psi T1} = \bar{q} \sigma_{\mu\nu}U P_{+}r
\langle \tau_{L} L_{\mu}L_{\nu}\rangle F_{\psi T1} &
\hspace{-0.75cm}\mathcal{O}_{\psi T5} = \bar{l} \sigma_{\mu\nu}U P_{12}\eta
\langle\tau_{L}L^{\mu}\rangle \langle U P_{21}U^{\dagger}L^{\nu}\rangle F_{\psi T5}\\
\mathcal{O}_{\psi T2} = \bar{q} \sigma_{\mu\nu}U P_{-}r
\langle \tau_{L} L_{\mu}L_{\nu}\rangle F_{\psi T2} &
\hspace{-0.75cm}\mathcal{O}_{\psi T6} = \bar{l} \sigma_{\mu\nu}U P_{-}\eta
\langle \tau_{L} L_{\mu}L_{\nu}\rangle F_{\psi T6} \\
\mathcal{O}_{\psi T3} = \bar{q} \sigma_{\mu\nu}U P_{12}r
\langle\tau_{L}L^{\mu}\rangle \langle U P_{21}U^{\dagger}L^{\nu}\rangle F_{\psi T3}\\
\mathcal{O}_{\psi T4} = \bar{q} \sigma_{\mu\nu}U P_{21}r
\langle\tau_{L}L^{\mu}\rangle \langle U P_{12}U^{\dagger}L^{\nu}\rangle F_{\psi T4}\\
\rule{0cm}{0.8cm}
\mathcal{O}_{\psi T7} = \bar{q} \sigma_{\mu\nu}U P_{+}r 
\langle\tau_{L}L^{\mu}\rangle \left(\partial^{\nu}\tfrac{h}{v}\right) F_{\psi T7} &
\hspace{-0.75cm}\mathcal{O}_{\psi T11} = \bar{l} \sigma_{\mu\nu}U P_{-}\eta 
\langle\tau_{L}L^{\mu}\rangle \left(\partial^{\nu}\tfrac{h}{v}\right) F_{\psi T11}\\
\mathcal{O}_{\psi T8} = \bar{q} \sigma_{\mu\nu}U P_{-}r 
\langle\tau_{L}L^{\mu}\rangle \left(\partial^{\nu}\tfrac{h}{v}\right) F_{\psi T8} &
\hspace{-0.75cm}\mathcal{O}_{\psi T12} = \bar{l}\sigma_{\mu\nu} U P_{12}\eta 
\langle U P_{21}U^{\dagger}L^{\mu}\rangle\left(\partial^{\nu}\tfrac{h}{v}\right)
F_{\psi T12}\\
\mathcal{O}_{\psi T9} = \bar{q}\sigma_{\mu\nu} U P_{21}r 
\langle U P_{12}U^{\dagger}L^{\mu}\rangle\left(\partial^{\nu}\tfrac{h}{v}\right)
F_{\psi T9}\\
\mathcal{O}_{\psi T10} = \bar{q}\sigma_{\mu\nu} U P_{12}r 
\langle U P_{21}U^{\dagger}L^{\mu}\rangle\left(\partial^{\nu}\tfrac{h}{v}\right)
F_{\psi T10}\\
\end{array}
\end{equation}
Here we have used
\begin{equation}\label{fstuh}
F_{\psi S(T)i}\equiv F_{\psi S(T)i}(h)=
1+\sum^\infty_{n=1}f_{\psi S(T)i,n}\left(\frac{h}{v}\right)^n
\end{equation}

For completeness, we also quote the terms of the form $\psi^{2}UhX$:
\begin{equation}
\begin{array}{ll}
\label{psi2uhx}
\mathcal{O}_{\psi X1} = \bar{q} \sigma_{\mu\nu}U P_{+}r B^{\mu\nu}
F_{\psi X1} \quad& \mathcal{O}_{\psi X5} = 
\bar{q} \sigma_{\mu\nu}U P_{12}r \langle U P_{21} U^{\dagger} W^{\mu\nu}\rangle
F_{\psi X5} \\
\mathcal{O}_{\psi X2} = \bar{q} \sigma_{\mu\nu}U P_{-}r B^{\mu\nu}
F_{\psi X2} & \mathcal{O}_{\psi X6} = 
\bar{q} \sigma_{\mu\nu}U P_{21}r \langle U P_{12} U^{\dagger} W^{\mu\nu}\rangle
F_{\psi X6}\\
\mathcal{O}_{\psi X3} = \bar{q} \sigma_{\mu\nu}U P_{+}r 
\langle \tau_{L} W^{\mu\nu}\rangle F_{\psi X3} & 
\mathcal{O}_{\psi X7} =\bar{q} \sigma_{\mu\nu}G^{\mu\nu}U P_{+}r
F_{\psi X7} \\
\mathcal{O}_{\psi X4} = \bar{q} \sigma_{\mu\nu}U P_{-}r 
\langle \tau_{L} W^{\mu\nu}\rangle F_{\psi X4} & 
\mathcal{O}_{\psi X8} =\bar{q} \sigma_{\mu\nu}G^{\mu\nu}U P_{-}r
F_{\psi X8} \\
\rule{0cm}{0.8cm}
\mathcal{O}_{\psi X9} = \bar{l} \sigma_{\mu\nu}U P_{-}\eta B^{\mu\nu}
F_{\psi X9} & \mathcal{O}_{\psi X10} = 
\bar{l} \sigma_{\mu\nu}U P_{-}\eta \langle \tau_{L} W^{\mu\nu}\rangle
F_{\psi X10} \\
\mathcal{O}_{\psi X11} = \bar{l} \sigma_{\mu\nu}U P_{12}\eta 
\langle U P_{21} U^{\dagger} W^{\mu\nu}\rangle F_{\psi X11} &
\end{array}
\end{equation}
where
\begin{equation}\label{fpsixuh}
F_{\psi Xi}\equiv F_{\psi Xi}(h)=
1+\sum^\infty_{n=1}f_{\psi Xi,n}\left(\frac{h}{v}\right)^n
\end{equation}

The operators $\psi^2 UhX$ are not required as NLO counterterms,
since the one-loop diagrams inducing these structures in the effective theory 
are finite.  
These operators are expected to contribute at NNLO.
Also the tensor operators in (\ref{psi2uhd2t}) are not generated as one-loop
counterterms. The genuine counterterms in the class $\psi^2 UhD^2$ are
then those with the scalar fermion currents given in (\ref{psi2uhd2s}).

\subsection{\boldmath $\psi^4Uh$ terms}

The 4-fermion operators of the class $\psi^4U$ have been
listed in \cite{Buchalla:2012qq}. Since no derivatives are involved,
the generalization to the case including the $h$ field simply amounts
to a multiplication of each of these operators with a general function
\begin{equation}\label{f4psiuh}
F_{4\psi i}(h)=1+\sum^\infty_{n=1}f_{4\psi i,n}\left(\frac{h}{v}\right)^n
\end{equation}
The operators in the class $\psi^4Uh$ are then given by
\begin{equation}\label{4psiuh}
{\cal O}_{4\psi Uh,i} = {\cal O}_{4\psi U,i}\, F_{4\psi i}(h)
\end{equation}
where ${\cal O}_{4\psi U,i}$ are the 4-fermion operators listed in
Section 4.5 of \cite{Buchalla:2012qq}.

Not all of these operators need actually appear as counterterms
at one loop. While for instance operators of the form 
$\bar\psi_L U\psi_R\, \bar\psi_L U\psi_R F(h)$ are required as
counterterms, the operators  
$\bar\psi_L \gamma^\mu\psi_L\, \bar\psi_L \gamma_\mu\psi_L F(h)$ are not.
Still the latter could arise as finite contributions at NLO through the 
tree-level exchange of TeV-scale resonances.

\subsection{\boldmath $X^3Uh$ terms}

The operators $X^3$, built from 3 factors
of field-strength tensors, are not required as counterterms at 
next-to-leading order. 
There are only four operators of this type 
\cite{Grzadkowski:2010es,Buchmuller:1985jz}
\begin{equation}\label{x3g}
{\cal O}_{X1}=f^{ABC} G^{A\nu}_\mu G^{B\rho}_\nu G^{C\mu}_\rho\, ,\qquad
{\cal O}_{X2}=f^{ABC} \tilde G^{A\nu}_\mu G^{B\rho}_\nu G^{C\mu}_\rho
\end{equation}
\begin{equation}\label{x3w}                                               
{\cal O}_{X3}=\varepsilon^{abc} W^{a\nu}_\mu W^{b\rho}_\nu W^{c\mu}_\rho\, ,\qquad    
{\cal O}_{X4}=\varepsilon^{abc} \tilde W^{a\nu}_\mu W^{b\rho}_\nu W^{c\mu}_\rho
\end{equation}
where $f^{ABC}$ and $\varepsilon^{abc}$ are the structure constants
of colour $SU(3)$ and weak $SU(2)$, respectively.
They are dimension-6 operators and therefore suppressed by 
two powers of the heavy mass scale $\Lambda$. A loop suppression
brings the coefficients further down to the NNLO level
${\cal O}(v^4/\Lambda^4)$ \cite{Appelquist:1993ka,Arzt:1994gp,Manohar:2013rga}
(see \cite{Buchalla:2012qq} for additional comments).
Similar arguments hold for the entire class of terms $X^3 Uh$,
that is including Goldstone and Higgs fields, which we do not consider
further here.

\section{Partial compositeness and the linear realization}
\label{sec:xicount}

The power-counting formula we have derived and applied in the preceding 
sections assumed that the scale of electroweak symmetry breaking $4\pi v$ 
and the cut-off scale $4\pi f$ were of comparable size. 
This situation includes nondecoupling 
scenarios, where there is only one relevant scale $v$ and the composite Higgs 
plays the role of a pseudo-Goldstone boson. In these scenarios, the full 
unitarization of amplitudes ({\it{e.g.}} in $W_LW_L$ scattering) is taken care 
of by states at the TeV scale. On the opposite end, $v/f\to 0$, we have 
the Standard Model Higgs, which alone unitarizes the physical amplitudes 
due to the renormalizability of its interactions. Between these two 
pictures, there is a continuum of possibilities where heavy resonances and a
light Higgs together render the theory unitary. 
In order to cover the transition between the pure nondecoupling case
(TeV-scale new states) and the Standard-Model scenario (infinitely heavy new 
states), the scales $f$ and $v$ should be distinguished. Theories with vacuum 
misalignment~\cite{Kaplan:1983fs,Dugan:1984hq}, for instance,
are examples of how this splitting of scales can be dynamically realized. 
The vacuum-tilting parameter
\begin{align}
\xi=\frac{v^2}{f^2}
\end{align} 
therefore gauges the degree of $h$-compositeness or, equivalently, the degree 
of decoupling of the theory: $\xi=1$ corresponds to purely nondecoupling 
scenarios, while $\xi\to0$ is the decoupling limit, {\it i.e.} the
Standard-Model case.

The relation between the two limits, $\xi=1$ and $\xi\ll 1$, can be made
more explicit. Since for $\xi\to 0$ the theory reduces to the renormalizable
Standard Model, with a linearly transforming Higgs doublet $\phi$, the 
effective Lagrangian can be organized for small $\xi$ in terms of operators
of increasing canonical dimension $d$. The coefficients of these
operators then scale as $\xi^{(d-4)/2}$.\footnote{Further small factors 
such as couplings or powers of $1/4\pi$, arising e.g. from resonance masses 
$M_R\sim 4\pi f$, will be ignored in the present context. The resulting 
suppression of particular coefficients can be separately addressed.}
This corresponds to the usual framework, of which the terms up to $d=6$ have 
been classified in \cite{Grzadkowski:2010es,Buchmuller:1985jz}. 
The restriction $\xi\ll 1$ may be relaxed by considering $\xi$ as a quantity 
of ${\cal O}(1)$. Then the effective theory in powers of $\xi$ has to be
reorganized in terms of the chiral Lagrangian. This effectively resums
the series in $\xi$, replacing it by a loop expansion. As a consequence
of the reorganization there is no one-to-one correspondence between the terms 
classified as NLO in the two scenarios, $\xi={\cal O}(1)$ and $\xi\ll 1$.
It also implies that (for most operator classes) the chiral Lagrangian
formulation is more general than the effective theory based on canonical
dimension, as explained in more detail below.

We may rewrite the dimension-6 operators from \cite{Grzadkowski:2010es},
whose coefficients count as ${\cal O}(\xi)$, in polar coordinates
for the Higgs field, using
\begin{equation}\label{phiuh}
\phi =(v+h)\,U\left(\begin{array}{c}0\\1\end{array}\right),
\qquad \tilde\phi =
(v+h)\,U\left(\begin{array}{c}1\\0\end{array}\right)
\end{equation}
The resulting terms can be matched to some of the operators in the chiral 
Lagrangian. The coefficients of those operators are then seen to start at
${\cal O}(\xi)$ in the small-$\xi$ limit. Higher powers of $\xi$ are 
always present in the expansion of these coefficients. This is because
additional factors of $\phi^\dagger \phi=(v+h)^2$, multiplying a given
operator, lead to higher-dimensional operators that map onto the same
operator in the chiral Lagrangian. Operators in the chiral Lagrangian that
cannot be obtained from the dimension-6 basis of \cite{Grzadkowski:2010es}
derive from operators of dimension $d > 6$. Their coefficients then count
as ${\cal O}(\xi^{(d-4)/2})$ in the small-$\xi$ expansion.

We illustrate this for the dimension-6 operators in the class 
$\psi^2\phi^2 D$ of \cite{Grzadkowski:2010es}. They have the form  
\begin{align}\label{opsivxi}
\bar q\gamma^\mu q\, \phi^\dagger i\overleftrightarrow{D}_\mu \phi &= 
2 (v+h)^2\, \bar q\gamma^\mu q\, \langle\tau_L L_\mu\rangle \\
\bar q\gamma^\mu T^a q\, \phi^\dagger i\overleftrightarrow{D}_\mu T^a\phi &=   
-\frac{1}{2} (v+h)^2\, \bar q\gamma^\mu L_\mu q \\
\bar u\gamma^\mu d\, \tilde\phi^\dagger i D_\mu \phi &=  
-(v+h)^2\, \bar u\gamma^\mu d\, \langle L_\mu UP_{21} U^\dagger\rangle 
\end{align}
with similar relations for the remaining operators. Recalling that 
\begin{equation}\label{qlqv332}
-\bar q \gamma^\mu L_\mu q F(h) = 
{\cal O}_{\psi V3} + {\cal O}^\dagger_{\psi V3} + 2 {\cal O}_{\psi V2} 
\end{equation}
we find that all operators in (\ref{psi2uhd}) are generated. Their
coefficients thus count as ${\cal O}(\xi)$. If we had used 
$\bar q \gamma^\mu L_\mu q F$ as a basis element instead of, say,
${\cal O}_{\psi V2}$, the operator ${\cal O}_{\psi V3}$ would not be
generated with an ${\cal O}(\xi)$ coefficient, but could only arise at
${\cal O}(\xi^2)$. This shows that the order in $\xi$ of the coefficients
in the chiral Lagrangian is in general basis dependent.

Mapping the entire dimension-6 basis of \cite{Grzadkowski:2010es} onto
the chiral Lagrangian, leads to the following list of chiral operators
with ${\cal O}(\xi)$ coefficients: 
\begin{align}\label{ochixi}
X^2 Uh,\, XUhD^2 &:\quad 
{\cal O}_{Xhi}, \quad i=1,\ldots, 6;\quad {\cal O}_{XU1}, {\cal O}_{XU4}\\
\psi^2 UhD &:\quad {\cal O}_{\psi Vi}, \quad i=1,\ldots, 10\\
\psi^4 Uh &:\quad \textrm{all 4-fermion operators without $U$-fields}
\end{align}

In these classes the chiral basis is more general than its dimension-6
counterpart: Not all chiral operators are generated from the dimension-6
basis, only terms up to second order in $h$ appear, and some of the
coefficients are correlated. The operators of classes $\phi^6$, $\phi^4 D^2$
and $\psi^2 \phi^3$ in \cite{Grzadkowski:2010es} contribute ${\cal O}(\xi)$
corrections to leading-order terms in the chiral 
Lagrangian.\footnote{One finds a direct correspondence between operators 
with the exception of the operator 
$(\phi^{\dagger}\phi)\Box(\phi^{\dagger}\phi)$, which in the chiral 
Lagrangian can be reabsorbed in terms of leading order coefficients
(see Appendix~\ref{sec:appllo}).}
The operators $X^3$ and $\psi^2 X\phi$ have ${\cal O}(\xi)$ coefficients,
but appear only at NNLO.

The remaining NLO operators in our basis for the chiral Lagrangian
have coefficients of higher order in $\xi$. 
For a complete classification of the various orders in $\xi$,
the lists of higher-dimensional operators in the Standard Model
would have to be worked out systematically beyond the dimension-6 level.
Since such lists are not yet available, we will content ourselves with 
commenting on a few typical cases. An important example is given by the
terms of class $UhD^4$ in Section \ref{subsec:uhd4}. 
The lowest-dimension, nonredundant 
operators that can generate them are operators in the pure-Higgs sector 
with four derivatives. The three independent terms in this class
are the dimension-8 operators  
\begin{equation}\label{phidim8}
D_\mu \phi^\dagger D^\mu \phi\, D_\nu \phi^\dagger D^\nu \phi,\quad
D_\mu \phi^\dagger D_\nu \phi\, D^\mu \phi^\dagger D^\nu \phi,\quad
D_\mu \phi^\dagger D_\nu \phi\, D^\nu \phi^\dagger D^\mu \phi
\end{equation}
Rewriting those in polar coordinates using (\ref{phiuh}), one finds that
all CP even operators ${\cal O}_{Di}$, $i=1,\ldots, 11$ are generated
with the exception of ${\cal O}_{D3}$. We conclude that these 10 operators
have coefficients starting at ${\cal O}(\xi^2)$.

Another example is given by the 4-fermion operators $\psi^4 Uh$
that explicitly include $U$ fields, such as terms of the form
$\bar\psi_L U\psi_R\, \bar\psi_L U\psi_R\, F(h)$. This term can only come from
a dimension-8 operator and thus also counts as ${\cal O}(\xi^2)$.

The comparison between the chiral Lagrangian discussed in this work
and the usual expansion in terms of canonical dimension, with a
linearly transforming Higgs doublet, is summarized in Table \ref{tab:dim6chi}.
\begin{table}[t]
\begin{center}
\small
\begin{tabular}{|l||c|c||c|c|c||c|c||c|c|}
\hline
${\cal L}_\chi$: & LO & LO & $X^2 Uh$ & $\psi^2 UhD$ & $\psi^4 Uh$ & 
$UhD^4$ & $\psi^2 UhD^2$ & NNLO & NNLO \\
 & & & $XUhD^2$ & & & & & & \\
\hline
${\cal L}_{BW}$: & $\phi^6$ & $\psi^2\phi^3$ & $X^2\phi^2$ & 
$\psi^2\phi^2 D$ & $\psi^4$ & NNLO & NNLO & $X^3$ & $\psi^2 X\phi$ \\
 & $\phi^4 D^2$ & & & & & & & & \\
\hline
\end{tabular}
\end{center}
\caption{\label{tab:dim6chi}
Correspondence between classes of NLO operators in the loop expansion of the
chiral Lagrangian (present work, first row) and the $1/f$ expansion 
of the effective Lagrangian based on canonical dimension 
(\cite{Grzadkowski:2010es,Buchmuller:1985jz}, second row).}
\end{table}

A special case of the small $\xi$ limit is the so-called Strongly-Interacting 
Light Higgs (SILH) model~\cite{Giudice:2007fh}, which considers a scenario 
where a composite scalar doublet $\phi$ gets nonstandard interactions, 
driven by a subset of $d=6$ operators. With the identification 
in~(\ref{phiuh}), the SILH Lagrangian can be rewritten in terms of the $U$ 
and $h$ fields and shown to correspond to a specific choice of the EFT 
coefficients. This exercise shows that:
\begin{itemize}
\item All the bosonic CP-even operators of 
Sections~\ref{subsec:uhd4} and \ref{subsec:xuhd}, to linear order in $\xi$, 
are present in the SILH Lagrangian with independent coefficients. 
\item Some of the SILH operators renormalize terms in the leading-order
Lagrangian (\ref{l4uh}).
\item SILH does not contain any explicit fermionic operator but includes the 
combinations $D^{\mu}W_{\mu\nu}$ and $\partial^\mu B_{\mu\nu}$, which can be 
reduced 
to fermionic operators by a straightforward application of the equations of 
motion for the gauge fields. This hypothesis of universality imposes strong 
constraints on the fermionic operators. In particular, the model does not 
contain NLO operators with tensor and scalar fermion bilinears, and only two 
independent combinations of fermionic vector currents are generated, namely 
$\sum_{f} Y_{f}\mathcal{O}_{\Psi Vf}$ and $2\mathcal{O}_{\Psi V2,8} + 
\mathcal{O}_{\Psi V3,9} + \mathcal{O}^{\dagger}_{\Psi V3,9}$. In turn, the 
four-fermion sector is constrained to three independent combinations of 
operators, coming from operators like 
$D^{\mu}W_{\mu\nu}D_{\lambda}W^{\lambda\nu}$ after using the equations of
motion.  
\item Two operators of the class $X^{3}$ are considered, which strictly 
speaking should be counted as next-to-next-to-leading order
$(1/16\pi^2) v^2/\Lambda^2$.  
\end{itemize}

\section{Comparison with previous literature}
\label{sec:complit}

Traditional effective field theory descriptions of EWSB with underlying strong 
dynamics have focused mainly on higgsless 
scenarios~\cite{Longhitano:1980tm,Appelquist:1980vg,Longhitano:1980iz}. 
While the idea of the Higgs as a composite pseudo-Goldstone, resulting from 
spontaneous breaking of either internal~\cite{Kaplan:1983fs} or space-time 
symmetries~\cite{Halyo:1991pc} was proposed much earlier, only recently 
these ideas have been cast in the language of EFTs. In most of the cases, 
effective operators have been constructed according to phenomenological needs, 
without aiming at completeness. 

To the best of our knowledge, the closest to a systematic classification of 
operators was done in~\cite{Alonso:2012px,Alonso:2012pz}, where the 
bosonic CP even sector and fermion bilinear operators were explored under
certain restrictions. In the following we list the main differences between 
\cite{Alonso:2012px,Alonso:2012pz} and the present paper: 
\begin{itemize}
\item The Higgs self-interacting operator 
${\cal{O}}_{D11}$ in (\ref{u0h4}) is not discussed in~\cite{Alonso:2012px}.
The CP odd bosonic operators have been omitted there, based on the assumption 
of CP invariance in the bosonic sector. Regarding the fermionic terms, 
Lorentz-vector bilinear operators in~\cite{Alonso:2012pz} are built only from
left-handed quark fields. If leptons and the right-handed fermions are also 
included, the basis gets enlarged from the 4 terms they consider to 13.
For the scalar and tensor bilinear sector, operators with derivatives on $h$ 
are not included. If one considers leptons and quarks, one finds 12 and 18 
operators, respectively, instead of the 4 and 6 listed 
in~\cite{Alonso:2012pz}. Finally, a discussion of four-fermion operators 
is absent.  
\item Comparing our basis to the set of 24 bosonic 
operators ${\cal P}_i$ in~\cite{Alonso:2012px}, we note that the 8 operators 
${\cal{P}}_4, {\cal{P}}_5, {\cal{P}}_{11}, {\cal{P}}_{12}, 
{\cal{P}}_{13}, {\cal{P}}_{14}, {\cal{P}}_{16}$ and ${\cal{P}}_{17}$ 
are redundant in the sense that they can be 
expressed as fermionic bilinear operators using the equations of motion 
for the gauge and $U$ fields. From the independent 16 operators in
\cite{Alonso:2012px}, the operators ${\cal{P}}_2$, ${\cal{P}}_3$,
${\cal{P}}_9$ are redundant in the absence of
$h$~\cite{Buchalla:2012qq,De Rujula:1991se,Nyffeler:1999ap,Grojean:2006nn}.
Therefore, they only appear with at least one power of $h$.
\item The assignment of powers of $\xi$ to
the various operators given in \cite{Alonso:2012px} is not in 
agreement with the discussion presented in Section \ref{sec:xicount}.
\end{itemize}
On a more general level, the major difference of 
\cite{Alonso:2012px,Alonso:2012pz} to our approach is that 
we rely on a consistent power-counting. This is not a mere technicality, but 
rather a fundamental issue in order to be able to organize the EFT expansion. 
In particular, without a power-counting one cannot even define a 
leading-order Lagrangian, let alone next-to-leading order corrections. 
One criticism one can raise against \cite{Alonso:2012px} is that they 
seem to use a naive dimensional power-counting, which is known to fail for 
strongly-coupled expansions. In particular, kinetic and mass terms for the 
gauge fields would have different power-counting dimensions, which is clearly 
inconsistent: both terms should instead be homogeneous and stand at the same 
order in the EFT expansion.

\section{Models of UV physics}
\label{sec:uvmod}

In this section we briefly discuss the SM effective Lagrangian
as a low-energy approximation of two simple models of physics
at higher energies. In the first part, we consider the MCHM5
model \cite{Agashe:2004rs,Contino:2006qr,Contino:2010rs} and show how 
the generic function $F_U(h)$ in (\ref{luh}) emerges in this case. 
In the second part we take a closer look at a specific UV-completion, 
based on the Higgs portal, 
and illustrate which operators of our NLO basis are generated. 

\subsection{MCHM5}

In the MCHM5, the four real Goldstone bosons $h_{a}$ are described 
by the vector pa\-ra\-me\-tri\-zing the coset $SO(5)/SO(4)$
\begin{equation}\label{eq:1}
  \vec\Sigma = \frac{\sin{\tfrac{|h|}{f}}}{|h|} 
 \left(h_{1},h_{2},h_{3},h_{4},|h| \cot{\tfrac{|h|}{f}}\right)^T,
\end{equation}
where $|h| = \sqrt{\sum_{a=1}^{4}(h_{a})^{2}}$. For the transition from the 
real 4-component vector $\vec{h}$ to the matrix $U$, we define
\begin{equation}\label{eq:2}
  (\langle h \rangle +h) U= i\sum^3_{a=1} h_a \sigma^a - h_4 \mathbf{1} = 
\begin{pmatrix}  -h_{4} + i h_{3} & h_{2}+i h_{1}\\-h_{2}+i h_{1} & 
                 -h_{4} - i h_{3}\end{pmatrix},
\end{equation}
where $(i\sigma^{a}, -\mathbf{1})$ defines a basis of $2 \times 2$ matrices 
with the Pauli-matrices $\sigma^{a}$, such that the 4 components $h_a$ are
related to the 4 real components $\phi_a$ of the Higgs doublet,
$\phi =(\phi_1 + i\phi_2,\phi_3 + i\phi_4)^T$ by a $SO(4)$ transformation. 
$\langle h \rangle$ is the vacuum expectation value of the 
scalar $|h|=\langle h\rangle + h$. An $SO(4)$ transformation that leaves 
$|h|$ invariant is then equivalent to an $SU(2)_{L}\times SU(2)_{R}$ 
transformation of the matrix $U$, defined in (\ref{uhglgr}). 
After expanding (\ref{uudef}) in terms of $\varphi$,
\begin{equation}\label{eq:3}
U = \cos{\tfrac{|\varphi|}{v}} + i \frac{\varphi^{a} \sigma^{a}}{|\varphi|}
\sin{\tfrac{|\varphi|}{v}},
\end{equation}
where 
$|\varphi|=\sqrt{(\varphi^{1})^{2}+(\varphi^{2})^{2}+(\varphi^{3})^{2}}$,
we find 
\renewcommand{\arraystretch}{1.5}
\begin{equation}\label{eq:4}
h_{a} = (\langle h \rangle+h)\, 
\frac{\varphi^{a}}{|\varphi|}\, \sin{\tfrac{|\varphi|}{v}},\,
a=1,2,3\quad \text{and} \quad 
h_{4} = -(\langle h \rangle+h) \cos{\tfrac{|\varphi|}{v}}
\end{equation}
\renewcommand{\arraystretch}{1}
Now we can write down the kinetic term of $\vec\Sigma$ 
in terms of $U$ and $h$,
\begin{equation}\label{eq:5}
 \frac{f^{2}}{2} D_{\mu}\vec\Sigma^T  D^{\mu}\vec\Sigma = 
\frac{1}{2} \partial_{\mu} h \partial^{\mu}h + 
\frac{f^{2}}{4} \langle D_{\mu}U^\dagger D^{\mu}U\rangle 
\sin^{2}{\left(\frac{\langle h \rangle+h}{f}\right)}
\end{equation}
By comparing this to (\ref{luh}) we can identify 
\begin{equation}\label{eq:6}
  \xi = \frac{v^{2}}{f^{2}}=\sin^{2}{\frac{\langle h\rangle }{f}}
\end{equation}
The coefficients $f_{U,n}$ in (\ref{luh}), for $n > 0$, are given by
\begin{equation}\label{eq:7}
 f_{U,n} = \frac{2}{n!}\begin{cases}(1-2\xi) (-4\xi)^{\tfrac{n}{2}-1}, & 
\text{for }n\text{ even}\\
  \sqrt{1-\xi}(-4\xi)^{\tfrac{n-1}{2}}, & \text{for }n\text{ odd}\end{cases}
\end{equation}
We see that each additional power of $(h/v)^2$ introduces a factor $\xi$. 
For $\xi\approx 1$ the odd powers of $h/v$ are suppressed in $F_U(h)$.

Finally, as an example of a NLO operator we may consider the
4-derivative term  $(D_{\mu}\vec\Sigma^T  D^{\mu}\vec\Sigma)^2$.
From (\ref{eq:5}) we see that in our basis it corresponds to a combination
of the operators ${\cal O}_{D1}$, ${\cal O}_{D7}$, ${\cal O}_{D11}$,
listed in Section~\ref{subsec:uhd4}.

\subsection{Higgs portal}
\label{subsec:higgsport}

As a specific model for a UV completion we consider the Higgs portal (see 
\citer{Schabinger:2005ei,Englert:2011yb}
and references therein). This model postulates the existence of a new, 
Standard-Model singlet scalar particle, which has allowed dimension-4 
couplings to the Higgs field. 
This interaction modifies the scalar potential of Eq. (\ref{luh}) to
\begin{equation}\label{eq:8}
V=-\frac{\mu_{s}^{2}}{2} |\phi_{s}|^{2} + 
\frac{\lambda_{s}}{4}|\phi_{s}|^{4}-\frac{\mu_{h}^{2}}{2} |\phi_{h}|^{2} + 
\frac{\lambda_{h}}{4}|\phi_{h}|^{4} + \frac{\eta}{2}|\phi_{s}|^{2}|\phi_{h}|^{2},
\end{equation}
where $\phi_{s}$ refers to the standard scalar doublet and $\phi_{h}$ 
denotes the hidden scalar. Both of them acquire a vacuum expectation value, 
which can be written as
\begin{equation}\label{eq:9}
\frac{v_{s}}{\sqrt{2}} = \sqrt{
\frac{\lambda_{h}\mu_{s}^{2}-\eta\mu^2_h}{\lambda_{s}\lambda_{h}-\eta^2}}, 
\qquad \frac{v_{h}}{\sqrt{2}} = \sqrt{
\frac{\lambda_{s}\mu_{h}^{2}-\eta\mu^2_s}{\lambda_{s}\lambda_{h}-\eta^2}}
\end{equation}
Expanding both scalars around their vacuum expectation value, i.e. 
$|\phi_{i}| = \tfrac{1}{\sqrt{2}}(v_{i}+h_{i})$, leads to a potential of 
the form
\begin{equation}\label{eq:10}
V=\frac{\lambda_s v_{s}^{2}}{4} h_{s}^{2} + 
\frac{\lambda_h v_{h}^{2}}{4} h_{h}^{2} + 
\frac{\eta}{2} v_{s}v_{h}h_{s}h_{h} + \mathcal{O}(h_{i}^{3})
\end{equation}
The transformation
\begin{equation}\label{eq:11}
\begin{pmatrix} H_{1}\\H_{2}\end{pmatrix} = \begin{pmatrix} \cos{\chi} & 
-\sin{\chi}\\\sin{\chi}&\cos{\chi}\end{pmatrix} 
\begin{pmatrix} h_{s}\\h_{h}\end{pmatrix}
\end{equation}
diagonalizes the mass matrix. The rotation angle $\chi$ is defined as
\begin{equation}\label{eq:12}
\tan{(2\chi)} = 
\frac{2\eta v_{s}v_{h}}{\lambda_h v_{h}^{2}-\lambda_s v_{s}^{2}}
\end{equation}
The masses of the physical states $H_{1}$ and $H_{2}$ are given by
\begin{equation}\label{eq:13}
M_{1,2}^{2} = \frac{1}{4}(\lambda_h v_{h}^{2}+\lambda_s v_{s}^{2}) \mp 
\frac{\lambda_h v_{h}^{2} - \lambda_s v_{s}^{2}}{4\cos{(2\chi)}}
\end{equation}
The Lagrangian relevant for the two scalars then reads
\begin{align}\label{eq:14}
\begin{aligned}
\mathcal{L}_{H} &= \frac{1}{2}\partial_{\mu}H_{1} \partial^{\mu}H_{1}+
\frac{1}{2}\partial_{\mu}H_{2} \partial^{\mu}H_{2} - V(H_{1},H_{2})\\
 &+\frac{v^{2}}{4}\langle D_{\mu}U^\dagger D^{\mu}U\rangle 
\left(1+\frac{2a_{1}}{v}H_{1}+\frac{2a_{2}}{v}H_{2} + 
\frac{b_{1}}{v^{2}}H_{1}^{2}+\frac{b_{12}}{v^{2}}H_{1}H_{2}+
\frac{b_{2}}{v^{2}}H_{2}^{2}\right)\\
  &-v \left(\bar{q}Y_{u}UP_{+}r + \bar{q}Y_{d}UP_{-}r +\bar{l}Y_{e}UP_{-}\eta +
\text{h.c.}\right)\left(1+\frac{c_{1}}{v}H_{1}+\frac{c_{2}}{v}H_{2}\right),
\end{aligned}
\end{align}
where 
\begin{align}\label{eq:15}
\begin{aligned}
V(H_{1},H_{2})&=\frac{1}{2} M_{1}^{2}H_{1}^{2}+ 
\frac{1}{2} M_{2}^{2}H_{2}^{2}-\lambda_{1}H_{1}^{3}-
\lambda_{2}H_{1}^{2}H_{2}-\lambda_{3}H_{1}H_{2}^{2}-\lambda_{4}H_{2}^{3}\\
 &- z_{1}H_{1}^{4}-z_{2}H_{1}^{3}H_{2}-z_{3}H_{1}^{2}H_{2}^{2}-z_{4}H_{1}H_{2}^{3}- 
  z_{5}H_{2}^{4}\\
\end{aligned}
\end{align}
The couplings $\lambda_{i}$ and $z_{i}$ depend on $\mu_{s}$, $\mu_{h}$, 
$\lambda_{s}$, $\lambda_{h}$ and $\eta$. With the  parameters of the
Higgs-portal model
\begin{equation}\label{eq:16}
a_{1} = \sqrt{b_{1}} = c_{1} = \cos{\chi}, \qquad a_{2}= \sqrt{b_{2}} = c_{2} = 
\sin{\chi}, \qquad b_{12} = 2 \sin{\chi}\cos{\chi},
\end{equation}
the theory is renormalizable and unitary. 
The scalar $H_{1}$ is now identified with the light 
scalar $h$ that was found at the LHC. $H_{2}$ is assumed to be 
heavy such that it can be integrated out. 
In doing so, we take its mass $M_2$ to be larger than all other energy scales
in the model, $M_2\gg v_h$, $v_s$, $M_1$.
In this limit the couplings $\lambda_s$, $\lambda_h$ and $\eta$ become
large. We will assume, however, that they still remain in a regime where
perturbation theory is a sufficiently reliable approximation.
$H_2$ can then be integrated out at tree level by solving its equation of
motion and inserting the solution into the Lagrangian (\ref{eq:14}).
The $H_2$-part of this Lagrangian can be written as
\begin{equation}\label{lh2}
{\cal L}_{H_2}=\frac{1}{2}H_2(-\partial^2-M^2_2)H_2
+J_1 H_2 + J_2 H^2_2 + J_3 H^3_2 + J_4 H^4_2
\end{equation}
where the $J_i$ can be read off from (\ref{eq:14}) and (\ref{eq:15}).
Making the dependence on $M_2$ explicit, the $J_i$ take the form
\begin{equation}\label{jjim2}
J_i\equiv M^2_2 J^0_i +\bar J_i
\end{equation}
where $J^0_i$ is a pure polynomial in $H_1\equiv h$.

The equation of motion for $H_2$ reads
\begin{equation}\label{eomh2}
(-\partial^2-M^2_2 + 2 J_2) H_2 +J_1 + 3 J_3 H^2_2 + 4 J_4 H^3_2 = 0
\end{equation}
It can be solved order by order in powers of $1/M^2_2$ by expanding
\begin{equation}\label{h2ser}
H_2= H^{(0)}_2 + H^{(1)}_2 + H^{(2)}_2 + \ldots,\qquad\quad 
H^{(l)}_2 = {\cal O}(1/M^{2l}_2)
\end{equation}
$H^{(0)}_2$ can be determined from the ${\cal O}(M^2_2)$ piece of 
(\ref{eomh2}) as an infinite series in powers of $h$
\begin{equation}\label{h20}
H^{(0)}_2 =\sum^\infty_{k=2} r_k h^k
\end{equation}
$H^{(1)}_2$ can then be obtained in terms of $H^{(0)}_2$, {\it etc.}.
Inserting the solution (\ref{h2ser}) of (\ref{eomh2}) back into 
(\ref{lh2}), (\ref{eq:14}), and expanding in $1/M^2_2$, one arrives at the
effective Lagrangian of the model with $H_2$ integrated out in the limit
described above. At leading order, ${\cal O}(1/M^0_2)$, the result has
the form of the electroweak chiral Lagrangian in (\ref{luh}), with the 
functions $F_U(h)$, $V(h)$ and $\sum_n \hat Y^{(n)}_f (h/v)^n$ given as
infinite series in $h$. For example,
\begin{equation}\label{fuh}
F_U(h)=2 a_1 \frac{h}{v} + (b_1 - a_1 a^2_2 (a_1 + a_2 v/v_h))
\frac{h^2}{v^2} + {\cal O}(h^3)
\end{equation}
 
Extending the derivation to the NLO terms of ${\cal O}(1/M^2_2)$
one finds
\begin{equation}\label{leffm2}
\mathcal{L}_{\text{eff}} =\mathcal{L}_{LO} + \Delta\mathcal{L}_{NLO}
+ \mathcal{O}(\tfrac{1}{M_{2}^{4}}), 
\end{equation}
where ($H^{(0)}_2\equiv H_0$)
\begin{equation}\label{dellnlo}
\Delta\mathcal{L}_{NLO}=\frac{\left[(-\partial^2+2\bar J_2)H_0 + \bar J_1 +
3\bar J_3 H^2_0 + 4\bar J_4 H^3_0\right]^2}{2 M^2_2 (1- 2 J^0_2-6 J^0_3 H_0
-12 J^0_4 H^2_0)}
\end{equation}
The effective Lagrangian $\Delta\mathcal{L}_{NLO}$ contains operators that 
modify the leading-order Lagrangian (\ref{luh}) as well as a subset of the 
next-to-leading operators of Section~\ref{sec:lhnlo}. 
In par\-ti\-cu\-lar, we have 
\begin{equation}\label{od1711}
{\cal O}_{D1}, {\cal O}_{D7}, {\cal O}_{D11};\quad
{\cal O}_{\psi S1}, {\cal O}_{\psi S2}, {\cal O}_{\psi S7},  
{\cal O}_{\psi S14}, {\cal O}_{\psi S15}, {\cal O}_{\psi S18},
\end{equation}
the hermitean conjugates of the ${\cal O}_{\psi Si}$ in (\ref{od1711}),
and 4-fermion operators coming from the square of the Yukawa bilinears
contained in $\bar J_1$. The 4-fermion operators that are generated have the 
same structure as those in the heavy-Higgs model discussed 
in \cite{Buchalla:2012qq}, which are
\begin{align}\label{eq:19}
\begin{aligned}
\mathcal{O}_{FY1}, \mathcal{O}_{FY3}, \mathcal{O}_{FY5}, \mathcal{O}_{FY7}, 
\mathcal{O}_{FY9}, \mathcal{O}_{FY10}, \mathcal{O}_{ST5}, \mathcal{O}_{ST9}, \\ 
\mathcal{O}_{LR1}, \mathcal{O}_{LR3}, \mathcal{O}_{LR8}, \mathcal{O}_{LR9}, 
\mathcal{O}_{LR10}, \mathcal{O}_{LR12}, \mathcal{O}_{LR17}, \mathcal{O}_{LR18}
\end{aligned}
\end{align}
and their hermitean conjugates, but they are now dressed with functions
$F_i(h/v)$.

This discussion shows explicitly how a subset of our NLO operators is 
generated in the Higgs-portal scenario. After integrating out the
heavy scalar $H_2$ in the non-decoupling limit $M_2\gg v_h$, $v_s$, $M_1$
the effective theory takes the form of a chiral Lagrangian.
In particular, even for $F_i(h/v)\to 1$, it is seen that operators of
canonical dimension 4 ($\mathcal{O}_{D1}$), 5 ($\mathcal{O}_{\psi Si}$)
and 6 (4-fermion terms) contribute at the same (next-to-leading) 
order $1/M^2_2$. This shows that the effective Lagrangian is not simply
organized in terms of canonical dimension.

\section{Conclusions}
\label{sec:concl}

The main results of this paper can be summarized as follows:
\begin{itemize}
\item
We formulate the most general effective field theory for the Standard Model
at the electroweak scale $v$, which includes a light scalar boson $h$,
singlet under the Standard-Model gauge group. The framework allows for the
possibility of dynamical electroweak symmetry breaking and a composite
nature of $h$.
\item
The leading-order Lagrangian is reviewed, emphasizing the assumptions
behind its construction.
\item
The resulting effective theory is nonrenormalizable in general,
with a cutoff at $\Lambda=4\pi v$ or above.
It takes the form of an electroweak chiral Lagrangian, generalized
to include the singlet scalar $h$. A power-counting analysis is used to
clarify the systematics of the effective theory beyond the leading order,
which is based on a loop expansion, rather than on the canonical dimension
of operators.
\item
The power-counting formula is used to identify the classes of
operators that are required as one-loop counterterms. The full set of
NLO operators is subsequently worked out.
\item
We discuss the relation between the chiral Lagrangian and the conventional
effective theory with a linearly transforming Higgs, based on operators
ordered by increasing canonical dimension.
We show that the usual dimension-6 Standard-Model Lagrangian and the SILH 
framework can be obtained as special cases from our scenario.
\item
To illustrate some important features of our formulation, we briefly
discuss two specific models within the context of the chiral Lagrangian,
the composite Higgs model based on $SO(5)/SO(4)$, and a simple, UV complete
model based on the Higgs portal mechanism.
\end{itemize}

The effective Lagrangian of the Standard Model we have 
constructed through next-to-leading order in the chiral expansion
can be used to analyse, in a model-independent way, new-physics effects in
processes at the TeV scale. Loop corrections can be systematically
included. Of particular interest will be the detailed investigation of
Higgs-boson properties, which should ultimately guide us to a deeper
understanding of electroweak symmetry breaking.

\appendix
\section{Leading-order effective Lagrangian}
\label{sec:appllo}

In this section we review the construction of the leading-order electroweak 
chiral Lagrangian of the Standard Model including a light Higgs 
singlet, ${\cal L}_{LO}(h)$, as given in (\ref{l4uh}) -- (\ref{fuvsum}). 
This Lagrangian is nonrenormalizable in general. 
It defines the starting point for the systematic power counting
on which the construction of the complete effective field theory is based.
This construction determines in particular the next-to-leading order operators,
which are the subject of the present work. Although the form of 
${\cal L}_{LO}(h)$ is known \cite{Contino:2010mh,Contino:2010rs}, 
it is worthwhile to discuss in detail the 
underlying assumptions. We will also emphasize a few features that
allow for simplifications in the final form of ${\cal L}_{LO}(h)$.

The effective Lagrangian is based on an expansion in powers of
$v^2/\Lambda^2$, where $v=246\,{\rm GeV}$ is the electroweak scale
and $\Lambda=4\pi v$ the scale of dynamical electroweak symmetry breaking.
To leading order the Lagrangian has to contain the unbroken, renormalizable
part of the Standard Model (\ref{lsm4}). It consists of dimension-4 terms,
which therefore scale as $v^4$, for processes at electroweak energies.
Electroweak symmetry breaking is introduced to leading order by
the Higgs sector Lagrangian ${\cal L}_{Uh}$ in (\ref{luh}).
The Goldstone sector provides masses to the $W$ and $Z$ bosons
through the $U$-field kinetic term $v^2\langle D_\mu U^\dagger D^\mu U\rangle$,
and to the fermions through the Yukawa interactions $v\bar\psi_L U\psi_R$.
Both scale as $v^4$, which identifies them as proper leading order terms, 
as it has to be the case. 
Note that the latter operators, and those in (\ref{lsm4}), have canonical
dimension two, three and four, respectively. This already implies that  
dimension alone is not the criterion by which the operators in the
effective Lagrangian are ordered.

We assume that the new strong dynamics respects the global
custodial symmetry $U\to g_L U g^\dagger_R$, $g_{L(R)} \in SU(2)_{L(R)}$,
to leading order. This singles out the term
$v^2\langle D_\mu U^\dagger D^\mu U\rangle$ for the pure Goldstone-boson
LO Lagrangian. The only further possible Goldstone term with two
derivatives that respects the SM gauge symmetry
\begin{equation}\label{b1term}
v^2 \langle U^\dagger D_\mu U T_3\rangle^2
\end{equation}
breaks the custodial symmetry and will be treated as a next-to-leading
order correction. This assumption is in line with the empirical fact that
there are no ${\cal O}(1)$ corrections to the electroweak $T$-parameter,
to which (\ref{b1term}) contributes. Custodial symmetry is still violated
at leading order by the Yukawa couplings and by weak hypercharge.
These effects introduce violations of custodial symmetry through
one-loop corrections, which also count as NLO terms.  

We next include the Higgs singlet $h$, considered as a light 
(pseudo-Goldstone) particle of the strong dynamics.
The field $h$ is strongly coupled to the Goldstone sector. This introduces
interactions with arbitrary powers $h^k$ that multiply the Goldstone
Lagrangian. Standard power counting (see e.g. \cite{Chivukula:2000mb} for
a review) then implies  
\begin{equation}\label{ghlambda}
{\cal L}=\frac{v^2}{4}\langle D_\mu U^\dagger D^\mu U\rangle\, 
\left( 1 + \sum^\infty_{k=1} f_k \left(\frac{g h}{\Lambda}\right)^k \right)
\end{equation}  
where the canonical form of the Goldstone kinetic term fixes the overall
normalization. In the present context $g$ stands for the generic 
Higgs-sector coupling. Additional derivatives scale as 
$\partial/\Lambda\sim v/\Lambda$ and are of higher order. For strong
coupling $g\approx 4\pi$ the new factor in (\ref{ghlambda}) then becomes a 
general function $1+F_U(h/v)$. Since $h$ scales as $v$, higher powers are
not suppressed.
This is similar to the field $U=\exp(2i\varphi^a T^a/v)$ containing all 
powers of $\varphi^a/v$.
However, since $h$ is a singlet, the coefficients $f_k$ are not further
restricted. The infinite number of $f_k$ reflects the composite nature
of the Higgs, whose internal structure cannot be fully described by
a finite number of terms. This limits the predictive power
of the effective theory to some extent. Nevertheless,
the theory still retains predictivity, since for processes with a given 
number of external $h$, and to a given loop order, only a finite number
of terms in the Lagrangian contributes. 

For the reasons just discussed, interactions with arbitrary powers
of $h/v$ are also included into the Yukawa terms in (\ref{luh}).

A kinetic term for $h$ has to be added to the Lagrangian, which
may be written as
\begin{equation}\label{lhkin}
{\cal L}_{h,kin}=\frac{1}{2} \partial_\mu h \partial^\mu h\, (1+F_h(h/v))
\end{equation}
Interactions described by a general function $F_h(h/v)$ have been added
to the pure kinetic term, following the same considerations that led to
the function $F_U(h/v)$ above. The Lagrangian in (\ref{lhkin}) is the
most general expression containing two derivatives and only $h$ fields.
It turns out, however, that the function $F_h$ can be removed by the
field redefinition
\begin{equation}\label{hhtilde}
\tilde h = \int_0^h \sqrt{1+F_h(s/v)}\, ds 
\end{equation} 
which transforms (\ref{lhkin}) into 
\begin{equation}\label{lhtilkin}                                             
{\cal L}_{h,kin}=\frac{1}{2} \partial_\mu \tilde h \partial^\mu \tilde h   
\end{equation}
Dropping the tilde, the kinetic term for $h$ takes the simple form
used in (\ref{luh}).

There are two further terms with two derivatives that can be built
from $U$ and $h$ fields. The first is the operator in (\ref{b1term}) 
multiplied by a function $F(h)$, the second is   
$\langle U^\dagger D_\mu U T_3\rangle\, \partial^\mu F(h)$.
Since they violate custodial symmetry in the sector built only from
$U$ and $h$ fields, we do not include them in the leading-order 
Lagrangian. As a contribution at next-to-leading order the second term
can be eliminated using the leading-order equations of motion, which
are given below. The first term remains as an operator at NLO. 

Lorentz invariant operators with $U$, $h$ and just a single derivative
cannot be formed. This leaves us to consider terms without derivatives,
constructed from $U$ and $h$ fields. Since $\langle U^\dagger U\rangle$
is a constant, and no other invariants can be obtained from $U$ alone,
the zero-derivative contribution in the scalar sector reduces to the
$h$-field potential $V(h)$. For the pseudo-Goldstone $h$ this potential
would be forbidden by shift symmetry, but it can be generated at the 
one-loop level (see \cite{Contino:2010rs} for a review).
Standard power counting for strong coupling, 
but including an overall loop factor $1/16\pi^2$, then gives
\begin{equation}\label{vhpow}
V(h)=\frac{1}{16\pi^2}\frac{\Lambda^4}{g^2} 
\sum_k f_{V,k} \left(\frac{gh}{\Lambda}\right)^k
=v^4 \sum_k f_{V,k} \left(\frac{h}{v}\right)^k
\end{equation} 
which again scales as a leading-order contribution.
This implies in particular that the physical Higgs mass is light, 
of order $v^2$, rather than $\Lambda^2$, as it would be the case for 
a typical strong-sector resonance. 
We remark that a linear term ($k=1$) in (\ref{vhpow}), which will arise for 
instance from tadpole diagrams, can always be eliminated by shifting the field 
$h$ and renormalizing other fields and parameters (such as $v$). Accordingly, 
$n\geq 2$ has been adopted for $V(h)$ in eq. (\ref{fuvsum}) of the main text.

In principle one might consider the coupling of powers of $h/v$ also
to the fermionic terms in (\ref{lsm4}), expressed through a generic function
$f(h)$ as
\begin{equation}\label{psikinh}
{\cal L_\psi}=\frac{i}{2} \bar\psi' \overleftrightarrow {\not\!\! D} \psi' \,
(1+f(h))^{-2} 
\end{equation}
The fermionic term has to be written here in its manifestly hermitean form, 
since the $h$-dependent factor prevents one from performing the usual 
simplification via integration by parts. A field redefinition
\begin{equation}\label{psiredef}
\psi' = \psi\, (1+f(h))
\end{equation}
brings (\ref{psikinh}) back to its conventional form 
${\cal L}_\psi=\bar\psi i\!\!\not\!\! D\psi$, up to a total derivative. 
This would redefine the Yukawa couplings $\hat Y^{(n)}$, but would leave the 
structure of (\ref{luh}) unchanged. The $h$-dependent pre\-factors 
in (\ref{psikinh}) can therefore be omitted. 

Finally, the possibility remains to dress the gauge-field terms
by Higgs-dependent functions, as in
\begin{equation}\label{xxfh}
\langle X_{\mu\nu} X^{\mu\nu}\rangle \, F_X(h)
\end{equation}
with $X_{\mu\nu}$ a field-strength tensor and $F_X(0)=0$. We assume that
the gauge field strengths are not strongly coupled to the Higgs sector.
The operators in (\ref{xxfh})  can arise at one loop with a coefficient 
$\sim 1/16\pi^2$, but not necessarily with any further suppression
in $1/\Lambda$. We therefore count them as terms of next-to-leading order.
This completes the explanation of the leading-order Lagrangian
in (\ref{lsm4}) and (\ref{luh}).

For convenience we quote the equations of motions implied by the 
leading-order Lagrangian (\ref{l4uh}) in the electroweak sector. 
They play an important role in simplifying the basis of operators at NLO 
and are given as follows:
\begin{equation}\label{eomb}
\partial^\mu B_{\mu\nu}=g' \left[Y_\psi\bar\psi\gamma_\nu\psi -\frac{i}{2} v^2
  \langle U^\dagger D_\nu UT_3\rangle (1+F_U(h))\right]
\end{equation}
\begin{equation}\label{eomw}
D^\mu W^a_{\mu\nu}=g \left[\bar\psi_L\gamma_\nu T^a\psi_L +\frac{i}{2} v^2
  \langle U^\dagger T^a D_\nu U\rangle (1+F_U(h))\right]
\end{equation}
\begin{align}\label{eomh}
& \partial^2 h+V'(h) = \frac{v^2}{4} \langle D_\mu U^\dagger D^\mu U\rangle
F'_U(h) \nonumber\\
& - \sum^\infty_{n=0} (n+1) \left( \bar q \hat Y^{(n+1)}_u UP_+r + 
\bar q \hat Y^{(n+1)}_d UP_-r + 
\bar l \hat Y^{(n+1)}_e UP_-\eta + {\rm h.c.}\right)
\, \left(\frac{h}{v}\right)^n
\end{align} 
\begin{align}\label{eomu}
&\frac{v}{2}\left[D_\mu\left( U^\dagger D^\mu U\, (1+F_U(h))\right)\right]_{ij} =
\nonumber\\
& \left[\hat Y+\sum^\infty_{n=1} \hat Y^{(n)}\left(\frac{h}{v}\right)^n\right]_{st}
(\bar\psi_{L,s} U)_j (P \psi_{R,t})_i - \left[\hat Y+
\sum^\infty_{n=1} \hat Y^{(n)}\left(\frac{h}{v}\right)^n\right]^\dagger_{ts}
(\bar \psi_{R,t} P)_j (U^\dagger \psi_{L,s})_i  \nonumber\\
& -\frac{1}{2}\delta_{ij}\left(
\left[\hat Y+\sum^\infty_{n=1} \hat Y^{(n)}\left(\frac{h}{v}\right)^n\right]_{st}
\bar \psi_{L,s} U P \psi_{R,t} - \left[\hat Y+
\sum^\infty_{n=1} \hat Y^{(n)}\left(\frac{h}{v}\right)^n\right]^\dagger_{ts}
\bar \psi_{R,t} P U^\dagger \psi_{L,s} \right)
\end{align}
Here $i$, $j$ are $SU(2)$ indices, $s$, $t$ are flavour indices, and the 
quantities $(\hat Y,\hat Y^{(n)}, P, \psi_L,\psi_R)$ are summed over 
$(\hat Y_u,\hat Y^{(n)}_u,P_+,q,r)$,
$(\hat Y_d,\hat Y^{(n)}_d,P_-,q,r)$ and $(\hat Y_e,\hat Y^{(n)}_e,P_-,l,\eta)$.
In a similar notation, the equations of motion for fermions
can be written as
\begin{align}\label{eompsi}
& i\!\not\!\! D\psi_L = 
v \left[\hat Y+\sum^\infty_{n=1} \hat Y^{(n)}\left(\frac{h}{v}\right)^n\right]
U P \psi_R \nonumber\\
& i\!\not\!\! D\psi_R = v 
\left[\hat Y+\sum^\infty_{n=1}\hat Y^{(n)}\left(\frac{h}{v}\right)^n\right]^\dagger
P U^\dagger \psi_L
\end{align}
where a summation over the appropriate terms on the right-hand sides
is understood.

\section*{Acknowledgements}

We thank Alejandro Celis for collaboration on the model discussed
in Sec. \ref{subsec:higgsport}. 
This work was performed in the context of the ERC Advanced Grant
project `FLAVOUR' (267104) and was supported in part by the 
DFG cluster of excellence `Origin and Structure of the Universe'.



\begin{thebibliography}{99}

\bibitem{Aad:2012tfa} 
  G.~Aad {\it et al.}  [ATLAS Collaboration],
  Phys.\ Lett.\ B {\bf 716}, 1 (2012)
  [arXiv:1207.7214 [hep-ex]].

\bibitem{Chatrchyan:2012ufa} 
  S.~Chatrchyan {\it et al.}  [CMS Collaboration],
  Phys.\ Lett.\ B {\bf 716}, 30 (2012)
  [arXiv:1207.7235 [hep-ex]].

\bibitem{ATLAS:2013sla} 
  ATLAS Collaboration,
  ATLAS-CONF-2013-034.

\bibitem{CMS:yva} 
  CMS Collaboration,
  CMS-PAS-HIG-13-005.

\bibitem{Aad:2013xqa} 
  G.~Aad {\it et al.}  [ ATLAS Collaboration],
  arXiv:1307.1432 [hep-ex].

\bibitem{Coleman:1969sm} 
  S.~R.~Coleman, J.~Wess and B.~Zumino,
  Phys.\ Rev.\  {\bf 177}, 2239 (1969):~
  C.~G.~Callan, Jr., S.~R.~Coleman, J.~Wess and B.~Zumino,
  Phys.\ Rev.\  {\bf 177}, 2247 (1969).

\bibitem{Feruglio:1992wf}
  F.~Feruglio,
  Int.\ J.\ Mod.\ Phys.\ A {\bf 8}, 4937 (1993)
  [hep-ph/9301281];
  J.~Bagger {\it et al.}, 
  Phys.\ Rev.\ D {\bf 49}, 1246 (1994)
  [hep-ph/9306256];
  V.~Koulovassilopoulos and R.~S.~Chivukula,
  Phys.\ Rev.\ D {\bf 50}, 3218 (1994)
  [hep-ph/9312317];
  C.~P.~Burgess, J.~Matias and M.~Pospelov,
  Int.\ J.\ Mod.\ Phys.\ A {\bf 17}, 1841 (2002)
  [hep-ph/9912459];
  B.~Grinstein and M.~Trott,
  Phys.\ Rev.\ D {\bf 76}, 073002 (2007)
  [arXiv:0704.1505 [hep-ph]].

\bibitem{Contino:2013kra} 
  R.~Contino, M.~Ghezzi, C.~Grojean, M.~M\"uhlleitner and M.~Spira,
  arXiv:1303.3876 [hep-ph].

\bibitem{Alonso:2012px} 
  R.~Alonso, M.~B.~Gavela, L.~Merlo, S.~Rigolin and J.~Yepes,
  Phys.\ Lett.\ B {\bf 722}, 330 (2013)
  [arXiv:1212.3305 [hep-ph]].

\bibitem{Contino:2010mh} 
  R.~Contino, C.~Grojean, M.~Moretti, F.~Piccinini and R.~Rattazzi,
  JHEP {\bf 1005}, 089 (2010)
  [arXiv:1002.1011 [hep-ph]].

\bibitem{Azatov:2012bz} 
  A.~Azatov, R.~Contino and J.~Galloway,
  JHEP {\bf 1204}, 127 (2012)
  [Erratum-ibid.\  {\bf 1304}, 140 (2013)]
  [arXiv:1202.3415 [hep-ph]].

\bibitem{Buchalla:2012qq} 
  G.~Buchalla and O.~Cata,
  JHEP {\bf 1207}, 101 (2012)
  [arXiv:1203.6510 [hep-ph]].
  
\bibitem{Longhitano:1980tm}
  A.~C.~Longhitano,
  Nucl.\ Phys.\  B {\bf 188}, 118 (1981).

\bibitem{Appelquist:1993ka}
  T.~Appelquist and G.~H.~Wu,
  Phys.\ Rev.\  D {\bf 48}, 3235 (1993)
  [arXiv:hep-ph/9304240].

\bibitem{Appelquist:1984rr}
  T.~Appelquist, M.~J.~Bowick, E.~Cohler and A.~I.~Hauser,
  Phys.\ Rev.\  D {\bf 31}, 1676 (1985).

\bibitem{Grzadkowski:2010es}
  B.~Grzadkowski, M.~Iskrzynski, M.~Misiak and J.~Rosiek,
  JHEP {\bf 1010}, 085 (2010)
  [arXiv:1008.4884 [hep-ph]].

\bibitem{Buchmuller:1985jz}
  W.~Buchm\"uller and D.~Wyler,
  Nucl.\ Phys.\  B {\bf 268}, 621 (1986).

\bibitem{Arzt:1994gp} 
  C.~Arzt, M.~B.~Einhorn and J.~Wudka,
  Nucl.\ Phys.\ B {\bf 433}, 41 (1995)
  [hep-ph/9405214].

\bibitem{Manohar:2013rga} 
  A.~V.~Manohar,
  arXiv:1305.3927 [hep-ph].

\bibitem{Kaplan:1983fs} 
  D.~B.~Kaplan and H.~Georgi,
  Phys.\ Lett.\ B {\bf 136}, 183 (1984).
  
\bibitem{Dugan:1984hq} 
  M.~J.~Dugan, H.~Georgi and D.~B.~Kaplan,
  Nucl.\ Phys.\ B {\bf 254}, 299 (1985).

\bibitem{Giudice:2007fh}
  G.~F.~Giudice, C.~Grojean, A.~Pomarol and R.~Rattazzi,
  JHEP {\bf 0706}, 045 (2007)
  [arXiv:hep-ph/0703164].

\bibitem{Appelquist:1980vg}
  T.~Appelquist and C.~W.~Bernard,
  Phys.\ Rev.\  D {\bf 22}, 200 (1980).

\bibitem{Longhitano:1980iz}
  A.~C.~Longhitano,
  Phys.\ Rev.\  D {\bf 22}, 1166 (1980).

\bibitem{Halyo:1991pc} 
  E.~Halyo,
  Mod.\ Phys.\ Lett.\ A {\bf 8}, 275 (1993).

\bibitem{Alonso:2012pz} 
  R.~Alonso, M.~B.~Gavela, L.~Merlo, S.~Rigolin and J.~Yepes,
  Phys.\ Rev.\ D {\bf 87}, 055019 (2013)
  [arXiv:1212.3307 [hep-ph]].



\bibitem{De Rujula:1991se}
  A.~De Rujula, M.~B.~Gavela, P.~Hernandez and E.~Masso,
  Nucl.\ Phys.\  B {\bf 384}, 3 (1992).

\bibitem{Nyffeler:1999ap}
  A.~Nyffeler and A.~Schenk,
  Phys.\ Rev.\  D {\bf 62}, 113006 (2000)
  [arXiv:hep-ph/9907294].

\bibitem{Grojean:2006nn}
  C.~Grojean, W.~Skiba and J.~Terning,
  Phys.\ Rev.\  D {\bf 73}, 075008 (2006)
  [arXiv:hep-ph/0602154].





\bibitem{Agashe:2004rs} 
  K.~Agashe, R.~Contino and A.~Pomarol,
  Nucl.\ Phys.\ B {\bf 719}, 165 (2005)
  [hep-ph/0412089].

\bibitem{Contino:2006qr} 
  R.~Contino, L.~Da Rold and A.~Pomarol,
  Phys.\ Rev.\ D {\bf 75}, 055014 (2007)
  [hep-ph/0612048].

\bibitem{Contino:2010rs}
  R.~Contino,
  arXiv:1005.4269 [hep-ph].

\bibitem{Schabinger:2005ei} 
  R.~Schabinger and J.~D.~Wells,
  Phys.\ Rev.\ D {\bf 72}, 093007 (2005)
  [hep-ph/0509209].

\bibitem{Patt:2006fw} 
  B.~Patt and F.~Wilczek,
  hep-ph/0605188.

\bibitem{Bowen:2007ia} 
  M.~Bowen, Y.~Cui and J.~D.~Wells,
  JHEP {\bf 0703}, 036 (2007)
  [hep-ph/0701035].

\bibitem{Englert:2011yb} 
  C.~Englert, T.~Plehn, D.~Zerwas and P.~M.~Zerwas,
  Phys.\ Lett.\ B {\bf 703}, 298 (2011)
  [arXiv:1106.3097 [hep-ph]].

\bibitem{Chivukula:2000mb}
  R.~S.~Chivukula,
  arXiv:hep-ph/0011264.


\end{thebibliography}
\end{document}